\documentclass{IEEEtran}
\pdfoutput=1
\usepackage[T1]{fontenc}
\usepackage{color}
\usepackage{float}
\usepackage{units}
\usepackage{amsmath}
\usepackage{amsthm}
\usepackage{amssymb}
\usepackage{stackrel}
\usepackage{graphicx}
\usepackage{tablefootnote}
\usepackage[unicode=true,
 bookmarks=true,bookmarksnumbered=true,bookmarksopen=true,bookmarksopenlevel=1,
 breaklinks=false,pdfborder={0 0 0},pdfborderstyle={},backref=false,colorlinks=false]
 {hyperref}
\hypersetup{pdftitle={Your Title},
 pdfauthor={Your Name},
 pdfpagelayout=OneColumn, pdfnewwindow=true, pdfstartview=XYZ, plainpages=false}

\makeatletter

\providecommand{\tabularnewline}{\\}
\floatstyle{ruled}
\newfloat{algorithm}{tbp}{loa}
\providecommand{\algorithmname}{Algorithm}
\floatname{algorithm}{\protect\algorithmname}

\theoremstyle{plain}
\newtheorem{thm}{\protect\theoremname}
\theoremstyle{definition}
\newtheorem{example}[thm]{\protect\examplename}
\theoremstyle{plain}
\newtheorem{lem}[thm]{\protect\lemmaname}

\usepackage[caption=false,font=normalsize,labelfont=sf,textfont=sf]{subfig}

\makeatother

\providecommand{\examplename}{Example}
\providecommand{\lemmaname}{Lemma}
\providecommand{\theoremname}{Theorem}

\begin{document}
\title{A Stochastic Particle Variational Bayesian Inference Inspired Deep-Unfolding
Network for Non-Convex Parameter Estimation}
\author{Zhixiang~Hu,~An~Liu,~\IEEEmembership{Senior Member,~IEEE},
and Minjian Zhao,~\IEEEmembership{Member,~IEEE}\thanks{Zhixiang Hu, An Liu and Minjian Zhao are with the College of Information
Science and Electronic Engineering, Zhejiang University, Hangzhou
310027, China, e-mail: \protect\href{http://anliu@zju.edu.cn}{anliu@zju.edu.cn}.}}
\maketitle
\begin{abstract}
Future wireless networks are envisioned to provide ubiquitous sensing
services, which also gives rise to a substantial demand for high-dimensional
non-convex parameter estimation, i.e., the associated likelihood function
is non-convex and contains numerous local optima. Variational Bayesian
inference (VBI) provides a powerful tool for modeling complex estimation
problems and reasoning with prior information, but poses a long-standing
challenge on computing intractable posteriori distributions. Most
existing variational methods generally rely on assumptions about specific
distribution families to derive closed-form solutions, and are difficult
to apply in high-dimensional, non-convex scenarios. Given these challenges,
firstly, we propose a parallel stochastic particle variational Bayesian
inference (PSPVBI) algorithm. Thanks to innovations such as particle
approximation, additional updates of particle positions, and parallel
stochastic successive convex approximation (PSSCA), PSPVBI can flexibly
drive particles to fit the posteriori distribution with acceptable
complexity, yielding high-precision estimates of the target parameters.
Furthermore, additional speedup can be obtained by deep-unfolding
(DU) the PSPVBI algorithm. Specifically, superior hyperparameters
are learned to dramatically reduce the number of algorithmic iterations.
In this PSPVBI-induced Deep-Unfolding Networks, some techniques related
to gradient computation, data sub-sampling, differentiable sampling,
and generalization ability are also employed to facilitate the practical
deployment. Finally, we apply the LPSPVBI to solve several important
parameter estimation problems in wireless sensing scenarios. Simulations
indicate that the LPSPVBI algorithm outperforms existing solutions.
\end{abstract}


\section{Introduction}

\IEEEPARstart{F}{uture} multi-functional wireless networks envisioned
by next-generation communication standards (e.g., 6G, Wi-Fi 7, and
Wi-Fi 8) are not only an extension of existing communication technology
but also are expected to provide various high-accuracy sensing services,
such as indoor positioning, activity recognition, radar sensing, Integrated
Sensing and Communications (ISAC) \cite{ISAC} and so on. These sensing
services drive the need for accurate estimation of multi-target parameters
such as distance, angle, and speed in complex wireless environment,
which often involves high-dimensional non-convex parameter estimation
problems. In this case, traditional parameter estimation methods such
as Maximum Likelihood (ML) esstimation and Maximum-a-Posteriori (MAP)
estimation can easily get trapped in a ``bad'' local optimum of
the corresponding high-dimensional non-convex objective function.
It is well known that Bayesian inference is more suitable for high-dimensional
non-convex parameter estimation because of its powerful ability in
exploiting prior information and modeling the uncertainty in parameters.
Moreover, Bayesian inference can provide the entire posteriori distribution
rather than a point estimate, so it is not easily trapped into local
optimums. However, it casts a long-standing challenge on deriving
intractable posteriori distributions. For most non-trivial models
and real-world applications, multiple integrals in a posteriori distribution
are impossible or very difficult to compute in closed form, so practitioners
must resort to approximate methods.

The methods of approximate posteriori can be broadly divided into
two categories, variational bayesian inference (VBI) \cite{VBI_Beal,VBI+IS}
and the Monte Carlo methods \cite{MCMC,SMC,WL_PASS}.

VBI can provide an analytical approximation to the posteriori distributions
by iteratively updating the variational probability distributions
to optimize Kullback-Leibler (KL) divergence. As a deterministic approximation,
VBI can handle high-dimensional parameter spaces by selecting suitable
optimization algorithms, so it is often scalable to large-scale data
sets and complex models \cite{SteinVI}. However, for the conventional
VBI, it is usually assumed that the distributions come from certain
special distribution families or satisfy conjugate conditions \cite{BlackboxVI}.
These assumptions are often not accurate enough and can be subjective.
Simple distribution sets may not be flexible enough to fit a true
posteriori distribution well, while more advanced choices pose difficulties
for the subsequent optimization tasks. For this reason, efficient
variational methods often need to be derived on a model-by-model basis,
causing is a major barrier for developing general purpose algorithms.

In another line of research, the Monte Carlo method approximates a
true posteriori by generating samples from distributions through sampling
techniques such as Markov chain Monte Carlo (MCMC) \cite{MCMC}, sequential
Monte Carlo (SMC) \cite{SMC}, importance sampling (IS) \cite{IS,WL_PASS},
etc. Therefore, the method can be more flexible in characterizing
prior and posteriori distributions, so it can be applied to a variety
of non-standard probability distributions and can be performed in
parallel. However, such methods are generally slow and difficult to
converge \cite{SteinVI}, and can also become inapplicable and inefficient
in high-dimensional problems.

Further combining the advantages of the two methods, particle VBI
(PVBI) came into being. The VBI framework ensures efficient probabilistic
inference in PVBI, and the sampling technique employed makes it scalable
to complex models. Some works have been done exploring this route.
In \cite{VBI+IS}, although discrete particles are used to fit the
variational posteriori, conjugate prior assumptions are still adopted,
which limits its scope of use. In \cite{PMD}, the method of particle
mirror descent (PMD) updates only the weights of the particles, but
not the positions of the particles. Although the other kernel-based
PMD proposed in this paper can change the particle positions by resampling,
the kernel-based method needs to deal with continuous multiple integrals,
which will increase the complexity significantly and may further lead
to weight degradation. In \cite{SteinVI}, with the theoretical guidance
of functional analysis, Stein Variational Gradient Descent (SVGD)
directly drives the particles to update their positions, and finally
approximates the posteriori distribution according to the statistics
of the aggregated particles. However, when there are many variables
and with a wide range of values, this approach becomes unacceptably
complex. An inappropriate kernel function can also result in degraded
performance of the SVGD. In order to further improve the approximate
efficiency of the particles and reduce the number of particles used
in PVBI, we decided to update both the weights and positions of the
particles to directly minimze the KL divergence. Such improved degree
of freedom can further enhance the performance and accelerate the
convergence speed.

In addition, for high-dimensional problems, the volume of sample space
increases exponentially, and effectively covering this large sample
space requires a large number of samples. The challenge of such a
dimensional disaster also requires PVBI to deal with. In \cite{BlackboxVI,SVI},
stochastic optimization is employed to optimize the variational objective
function in the VBI, where a fast noisy approximation of the gradient
is calculated from Monte Carlo samples from the variational distribution.
With certain conditions on the step-size schedule, the noisy estimates
of a gradient guarantee these algorithms can provably converge to
an optimum, which are often cheaper to compute than the true gradient.
On this basis, we refer to stochastic successive convex approximation
(SSCA) \cite{SSCA,PSSCA} and propose a novel stochastic particle
variational Bayesian inference (SPVBI) algorithm in \cite{SPVBI}
for the special case of multiband sensing problem. The algorithm takes
an average in both gradient and iterates (optimization variables)
with decreasing step sizes, and updates optimization variables alternatively
by solving a sequence of convex surrogate subproblems. In summary,
SPVBI algorithm combines variational Bayesian inference, Monte Carlo
sampling, stochastic optimization to give full play to their respective
advantages. However, the SPVBI in \cite{SPVBI} is specifically designed
for multiband sensing. Moreover, although the convergence can be guaranteed
by alternating optimization with the decreasing step sizes, the convergence
speed is still slow especially for high-dimensional problems.

Most practical parameter estimation applications require immediate
results in the case of low latency and limited complexity, which implies
that algorithms can afford only a very small number of iterations
(e.g., ten or fewer) \cite{Survey_DU1,Survey_DU2}. In order to accelerate
the convergence speed, it is necessary to carefully tune the hyperparameters
(e.g., step-size selection), which is heuristic and not reliable.
To solve this problem, a model-driven deep learning scheme, deep unfolding
(DU), has been proposed recently. The main idea of DU is to unfold
the iterative algorithm as a series of neural network layers with
some learnable parameters. Although there are currently some VBI-inspired
DU networks \cite{DUVBI,DUVBI_HZ,DUVBI_LY}, they are all based on
conventional VBI algorithms and are heavily constrained by distribution
assumptions and application scenarios, thus exhibiting obvious limitations.
In addition, for the process of unfolding SPVBI algorithm, a series
of challenges still need to be overcome, such as the feasibility of
network backpropagation, the generalization ability and so on.

In this paper, we propose a parallel stochastic particle variational
Bayesian inference (PSPVBI) algorithm, which can be viewed as an extension
of the SPVBI in \cite{SPVBI} from multi-band sensing to more general
parameter estimation scenes, and from alternating optimization to
parallel optimization to facilitate deep unfolding and convergence
acceleration. In addition, we deep-unfold PSPVBI to derive a learnable
PSPVBI (LPSPVBI) algorithm to further reduce the complexity. The main
contributions are summarized below.
\begin{itemize}
\item \textbf{A parallel stochastic particle VBI method and its convergence
proof. }The proposed PSPVBI algorithm combines the advantages of variational
inference and Monte Carlo sampling to approximate the posteriori probability
distribution, and can describe various prior and posteriori distributions
flexibly and efficiently, and can also be applied to high-dimensional
non-convex problems. Updating particle weights and particle positions
simultaneously reduces the number of required particles and speeds
up convergence. In addition, we introduce parallel stochastic successive
convex approximations (PSSCA) to deal with the tricky crux of Bayesian
inference regarding multiple integrals. Different from stochastic
VBI (SVBI) \cite{SVI}, we average the gradient at the same time as
the iterates, which makes the noisy estimates of gradient smooth continuously
with the iteration, and also provides convergence guarantee for the
algorithm.
\item \textbf{Learnable PSPVBI greatly accelerating convergence and reducing
computational complexity.} By fusing the PSPVBI algorithm that have
performance guarantees with tools from deep learning, key hyperparameters
in the principled algorithm can be learned, greatly accelerating convergence
and reducing overall algorithm complexity. Existing performance guarantees
for the original PSPVBI algorithm can apply verbatim to learned unfolded
networks and appropriate constraints can be imposed on the learned
parameters. The resulting unfolded algorithm is intuitive, interpretable,
and has low complexity and memory requirements, which is in stark
contrast to black-box NNs.
\item \textbf{Specific} \textbf{deep-unfolding} \textbf{algorithm design
for actual deployment.} We apply the LPSPVBI to solve several important
application problems on parameter estimation in sensing scenes. Simulation
results show that the proposed LPSPVBI algorithm achieves competitive
performance for both scenes. For deployment details, we also provided
some practical tricks. For example, HyperNet \cite{HyperNet,HyperNet2}
is used to improve the generalization ability of deep unfolding network,
sharing intermediate computation results and subsampling observed
data are used to reduce the cost of computing gradients, and deep
neural network (DNN) is used to fit the sampling process to make sampling
efficient and backpropagated.
\end{itemize}
The rest of the paper is organized as follows. The general problem
formulation is given in Section $\text{\mbox{II}}$, together with
two application examples. The PSPVBI algorithm and the convergence
analysis are presented in Section $\text{\mbox{III}}$. The learnable
PSPVBI algorithm is proposed in Section $\text{\mbox{IV}}$. In Section
$\text{\mbox{V}}$, we present numerical simulations and performance
analysis. Section $\text{\mbox{VI}}$ applies the LPSPVBI algorithm
to solve several important application problems. Finally, conclusions
are given in Section $\text{\mbox{VI}I}$.

Notations: $\delta(\cdot)$ denotes the Dirac\textquoteright s delta
function, $vec\left[\cdot\right]$ denotes the vectorization operation,
$\propto$ denotes the left is proportional to the right, $\mathop{{\rm Re}}(\cdot)$
denotes the real part operator, $\mathop{{\rm var}}(\cdot)$ denotes
the variance operator, and $\left\Vert \cdot\right\Vert $ denotes
the Euclidean norm. For a matrix $\mathbf{A}$, $\mathbf{A}^{T}$
, $\mathbf{A}^{H}$, $\mathbf{A}^{-1}$, represent a transpose, complex
conjugate transpose and inverse, respectively. $\mathbb{E}_{z}[\cdot]$
denotes the expectation operator with respect to the random vector
$z$. $\boldsymbol{D}_{KL}\left[q\left\Vert p\right.\right]$ denotes
the Kullback-Leibler (KL) divergence of the probability distributions
$q$ and $p$. $\mathcal{N}(\mu,\Sigma)$ and $\mathcal{CN}(\mu,\Sigma)$
denotes Gaussian and complex Gaussian distribution with mean $\mu$
and covariance matrix $\Sigma$.

\section{General Problem Formulation for Non-Convex Prameter Estimation\label{sec:Problem-Formulations}}

Consider the following general measurement model of non-convex parameter
estimation:
\begin{equation}
r_{n}=f_{n}\left(\boldsymbol{\theta}\right)+w_{n},\forall n=1,2,\ldots N,
\end{equation}
where $r_{n}$ denotes the $n$-th observed data, $N$ is the total
number of observations, and $w_{n}$ denotes an additive white Gaussian
noise (AWGN) following the distribution $\mathcal{CN}\left(0,\eta_{w}^{2}\right)$.
The variables to be estimated are denoted as $\boldsymbol{\theta}{\rm =}\left[\theta_{1},\ldots,\theta_{J}\right]^{T}$,
and $J$ is the number of variables. $f_{n}\left(\boldsymbol{\theta}\right)$
is defined as the measurement function reconstructed from the parameters
$\boldsymbol{\theta}$.

All observed data can be vectorized as $\boldsymbol{r}=\left[r_{1},r_{2},\ldots,r_{N}\right]^{T}$.
In the case of AWGN, the likelihood function of the measurement model
can be written as follow
\begin{equation}
p\left(\boldsymbol{r}|\boldsymbol{\theta}\right)=\prod\limits _{n=1}^{N}p(r_{n}|\boldsymbol{\theta})=\prod\limits _{n=1}^{N}\frac{1}{\sqrt{2\pi}\eta_{w}}exp\left(\frac{\left|r_{n}-f_{n}\left(\boldsymbol{\theta}\right)\right|^{2}}{-2\eta_{w}^{2}}\right).
\end{equation}

The key step in Bayesian inference is to obtain the posteriori distribution,
denoted by $p\left(\boldsymbol{\theta}|\boldsymbol{r}\right)$. Following
the Bayes\textquoteright{} theorem, the posteriori distribution of
the unknown parameters is $p\left(\boldsymbol{\theta}|\boldsymbol{r}\right)\propto p\left(\boldsymbol{r}|\boldsymbol{\theta}\right)p\left(\boldsymbol{\theta}\right)$,
where $p\left(\boldsymbol{\theta}\right)$ denotes the potential prior
distribution. For a variable $\theta_{j}$ of interest, the marginal
posteriori distribution is denoted as
\begin{equation}
p\left(\theta_{j}|\boldsymbol{r}\right)\propto\int p\left(\boldsymbol{r}|\boldsymbol{\theta}\right)p\left(\boldsymbol{\theta}\right)d\boldsymbol{\theta}_{\sim j},
\end{equation}
where $\boldsymbol{\theta}_{\sim j}$ represents all the other variables
except $\theta_{j}$. Generally, it is intractable to get a closed-form
solution for $p\left(\theta_{j}|\boldsymbol{r}\right)$ due to the
high dimensional integrals over $\boldsymbol{\theta}_{\sim j}$ and
complicated measurement functions $f_{n}\left(\boldsymbol{\theta}\right)$.

VBI casts the posteriori inference as an optimization problem, where
variational distributions $q\left(\boldsymbol{\theta}\right)$ are
adjusted to approximate the true posteriori distributions $p\left(\boldsymbol{\theta}|\boldsymbol{r}\right)$.
The closeness is measured with Kullback-Leibler divergence, which
is defined as
\begin{equation}
\begin{split}\boldsymbol{D}_{KL}\left[q|p\right] & {\rm =}\int q\left(\boldsymbol{\theta}\right)ln\frac{q\left(\boldsymbol{\theta}\right)}{p\left(\boldsymbol{\theta}|\boldsymbol{r}\right)}d\mathbf{\boldsymbol{\theta}}\\
 & {\rm =}\int q\left(\boldsymbol{\theta}\right)ln\frac{q\left(\boldsymbol{\theta}\right)p\left(\boldsymbol{r}\right)}{p\left(\boldsymbol{r}|\boldsymbol{\theta}\right)p\left(\boldsymbol{\theta}\right)}d\boldsymbol{\theta},
\end{split}
\end{equation}
where $q\left(\boldsymbol{\theta}\right)$ is assumed to be factorized
as \cite{VBI_Beal,meanfield} $q\left(\boldsymbol{\theta}\right)=\stackrel[j=1]{J}{\prod}q\left(\theta_{j}\right)$,
where $q\left(\theta_{j}\right)$ stands for the approximation to
the marginal posteriori $p\left(\theta_{j}|\boldsymbol{r}\right)$.

Considering that $p\left(\boldsymbol{r}\right)$ is a constant independent
of $q\left(\boldsymbol{\theta}\right)$, minimizing the KL divergence
is equivalent to solving the following optimization problem:
\begin{equation}
\begin{split}\mathcal{P}_{1}:\mathbf{\mathbf{\boldsymbol{\theta}}^{*}=}\mathrm{arg}\mathop{\mathrm{min}}\limits _{q}\quad & \int q\left(\boldsymbol{\theta}\right)ln\frac{q\left(\boldsymbol{\theta}\right)}{p\left(\boldsymbol{r}|\boldsymbol{\theta}\right)p\left(\boldsymbol{\theta}\right)}d\boldsymbol{\theta}\end{split}
.\label{eq:ori_VBI_problem}
\end{equation}

In the following, we give some important application examples of the
problem formulation in \eqref{eq:ori_VBI_problem}.
\begin{example}
\label{exa:WL_exam}(Wireless cooperative localization \cite{WL_PASS}).
Consider a two-dimensional static wireless network, wherein the location-aware
nodes are assumed to be uniformly distributed within the deployment
area, as illustrated in Fig. \ref{WL}. For a target node, there are
$N_{R}$ reference nodes in its measurement range $r_{s}$, whose
initial locations are assumed to be inaccurate, due to errors in coarse
acquisition. Moreover, the absence of anchors throughout the entire
area necessitates collaborative localization efforts among the nodes
themselves. The true (but unknown) position vectors of the target
node and the $i$-th reference node are denoted as $\boldsymbol{s}_{0}=[px_{0},py_{0}]^{T}$
and $\boldsymbol{s}_{i}=[px_{i},py_{i}]^{T}$, $\forall i=1,\ldots,N_{R}$,
respectively, while the coarse locations (with a precision $\boldsymbol{U}_{i}$,
$\forall i=0,1,\ldots,N_{R}$) are denoted as $\boldsymbol{\mu}_{i},\forall i=0,1,\ldots,N_{R}$.
In general, the true location $\boldsymbol{s}_{i}$, $\forall i=0,1,\ldots,N_{R}$,
can be modeled as Gaussian variables \cite{Gaussprior}, characterized
by a mean of $\boldsymbol{\mu}_{i}$ and precision $\boldsymbol{U}_{i}$,
$\boldsymbol{s}_{i}\sim\mathcal{N}\left(\boldsymbol{\mu}_{i},\boldsymbol{U}_{i}\right)$,
$\forall i=0,1,\ldots,N_{R}$, where we assume node coarse locations
and precisions are independent to each other \cite{precision1,precision2}.
The prior information $\boldsymbol{\mu}_{i}$ and $\boldsymbol{U}_{i}$
recorded by each node, together with the measurement $\boldsymbol{z}_{i},\forall i=1,\ldots,N_{R}$
between the target node and the reference nodes, are shared with the
target node for cooperative localization.
\begin{figure}[htbp]
\begin{centering}
\textsf{\includegraphics[scale=0.5]{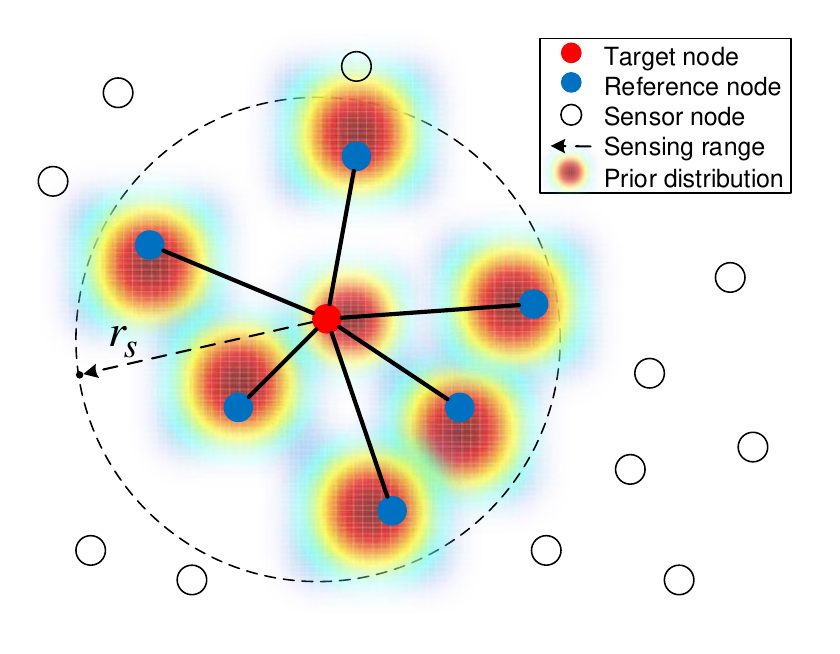}}
\par\end{centering}
\caption{\textsf{\label{WL}}Illustration of the network node deployment.}
\end{figure}

The measurement model can be given by
\begin{equation}
\boldsymbol{z}_{i}=h\left(\boldsymbol{s}_{i},\boldsymbol{s}_{0}\right)+\varepsilon_{i},\forall i=1,\ldots,N_{R},
\end{equation}
where $h\left(\boldsymbol{s}_{i},\boldsymbol{s}_{0}\right)$ denotes
the measurement function, and $\varepsilon_{i}$ represents the measurement
noise following the distribution $\mathcal{N}\left(0,U_{ni}\right)$,
where $U_{ni}$ is the precision parameter. Specifically, this example
adopt the received signal strength (RSS)-based measurement function,
which is
\begin{equation}
h\left(\boldsymbol{s}_{i},\boldsymbol{s}_{0}\right)=\phi_{ref}-10\lambda log_{10}\left\Vert \boldsymbol{s}_{i}-\boldsymbol{s}_{0}\right\Vert _{2},\label{eq:mea_fun_WL}
\end{equation}
where $\phi_{ref}$ denotes the reference power loss and $\lambda$
denotes the path loss exponent \cite{WL_PASS}. Let $\boldsymbol{c}=vec\left[\boldsymbol{s}_{i}\right]_{i=1}^{N_{R}}$
denote the reference node cluster of the target node, and $\boldsymbol{z}_{i}$
are stacked as $\boldsymbol{z}=vec\left[\boldsymbol{z}_{i}\right]_{i=1}^{N_{R}}$.
Under such a measurement model, the likelihood function can be written
as follow
\begin{equation}
p\left(\boldsymbol{z}|\boldsymbol{s}_{0},\boldsymbol{c}\right)=\stackrel[i=1]{N_{R}}{\prod}\frac{\left|U_{ni}\right|^{\frac{1}{2}}}{\sqrt{2\pi}}exp\left[-\frac{U_{ni}}{2}\left(\boldsymbol{z}_{i}-h\left(\boldsymbol{s}_{i},\boldsymbol{s}_{0}\right)\right)^{2}\right],
\end{equation}
and the posteriori distribution is cast as
\begin{equation}
\begin{split}p\left(\boldsymbol{s}_{i}|\boldsymbol{z}\right) & \propto\mathcal{N}\left(\boldsymbol{\mu}_{0},\boldsymbol{U}_{0}\right)\cdot\stackrel[i=1]{N_{R}}{\prod}\mathcal{N}\left(\boldsymbol{\mu}_{i},\boldsymbol{U}_{i}\right)\cdot\int\frac{\left|U_{ni}\right|^{\frac{1}{2}}}{\sqrt{2\pi}}\\
 & exp\left[-\frac{1}{2}U_{ni}\left(\boldsymbol{z}_{i}-h\left(\boldsymbol{s}_{i},\boldsymbol{s}_{0}\right)\right)^{2}\right]d\boldsymbol{s}_{i}.
\end{split}
\end{equation}

Due to the nonlinear measurement function \eqref{eq:mea_fun_WL} and
reference node location errors, the posteriori distribution of the
target node is non-convex and has many bad local optimums \cite{WL_PASS},
which complicates the determination of the optimal target node location.
\end{example}
\medskip{}

\begin{example}
\label{exa:MB_exam}(Multiband WiFi sensing \cite{SPVBI,wyb_MB}).
In the scene of multiband WiFi sensing, we consider a single-input
single-output (SISO) system that uses OFDM training signals over $M$
frequency subbands to estimate range between the mobile node and Wi-Fi
device. The discrete received signal model in frequency domain can
be formulated as \cite{wyb_MB}
\begin{equation}
\begin{split}r_{m}^{\left(n\right)} & =\sum\limits _{k=1}^{K}\alpha_{k}e^{j\beta_{k}}e^{-j2\pi\left(f_{c,m}+nf_{s,m}\right)\left(\tau_{k}+\delta_{m}\right)}e^{j\phi_{m}}+w_{m}^{\left(n\right)},\end{split}
\label{eq:original_model}
\end{equation}
where $m=1,2,\ldots M$ is the frequency band index, $N_{m}$ denotes
the number of subcarriers in each band, $n=0,1,\ldots N_{m}-1$ denotes
subcarrier index and $k=1,2,\ldots K$ denotes the $k$-th scattering
path. $\alpha_{k}e^{j\beta_{k}}$ is a complex scalar carrying the
amplitude and phase information of a scattering path, and $\tau_{k}$
is the time delay of the $k$-th path. $f_{c,m}$ and $f_{s,m}$ are
the initial frequency and subcarrier spacing of $m$-th frequency
band, respectively. In addition, due to the hardware imperfection,
the CSI measurements are superimposed with a random initial phase
$\phi_{m}$ and a timing synchronization error $\delta_{m}$. $w_{m}^{\left(n\right)}$
denotes an additive white Gaussian noise (AWGN) following the distribution
$\mathcal{CN}\left(0,\eta_{w}^{2}\right)$.
\begin{figure}[htbp]
\begin{centering}
\textsf{\includegraphics[scale=0.35]{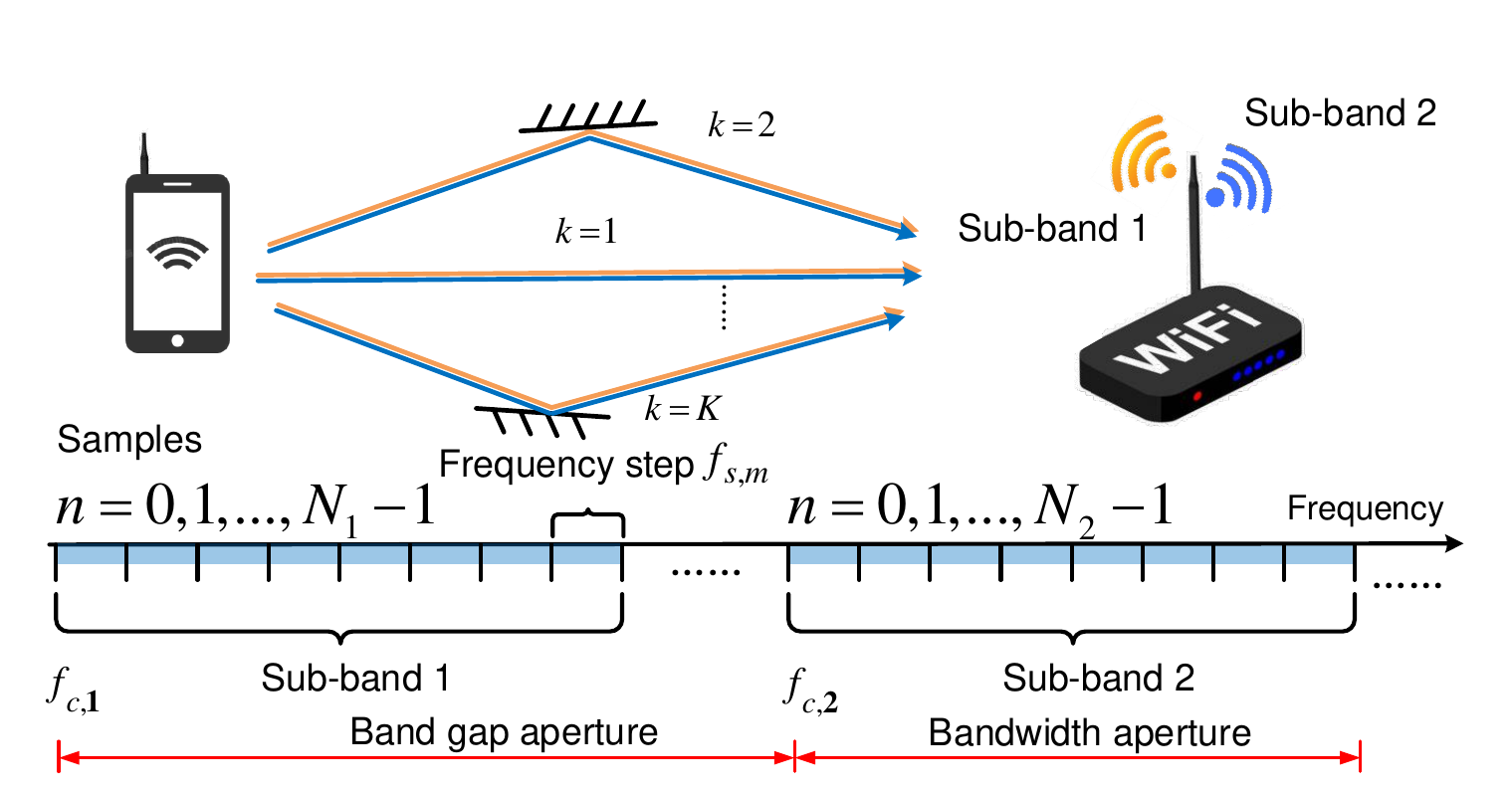}}
\par\end{centering}
\caption{\textsf{\label{MB}}Illustration of the multiband WiFi sensing.}
\end{figure}

In this case, the variables to be estimated are denoted as $\mathbf{\boldsymbol{\Lambda}_{ori}}{\rm =}\left[\alpha_{1},\ldots,\alpha_{K},\tau_{1},\ldots,\tau_{K},\beta_{1},\ldots,\beta_{K},\right.$
$\left.\phi_{1},\ldots,\phi_{M},\delta_{1},\ldots,\delta_{M}\right]^{T}$,
and frequency domain measurements of the received signal is vectorized
as $\boldsymbol{r}=\left[r_{1}^{\left(0\right)},r_{1}^{\left(1\right)},\ldots,r_{1}^{\left(N_{1}-1\right)},\right.$
$\left.\ldots,r_{M}^{\left(0\right)},r_{M}^{\left(1\right)},\ldots,r_{M}^{\left(N_{M}-1\right)}\right]^{T}$.
In the case of AWGN, the logarithmic likelihood function of the original
signal model \eqref{eq:original_model} can be written as follow
\begin{equation}
\begin{split}lnp\left(\boldsymbol{r}|\mathbf{\boldsymbol{\Lambda}_{ori}}\right)=ln & \prod\limits _{m=1}^{M}\prod\limits _{n=0}^{N_{m}-1}p(r_{m}^{\left(n\right)}|\mathbf{\boldsymbol{\Lambda}_{ori}})\\
=MN_{m}ln\frac{1}{\sqrt{2\pi}\eta_{w}} & -\sum_{m=1}^{M}\sum_{n=0}^{N_{m}-1}\frac{1}{2\eta_{w}^{2}}\left|r_{m}^{\left(n\right)}-s_{m}^{\left(n\right)}\left(\mathbf{\boldsymbol{\Lambda}_{ori}}\right)\right|^{2},
\end{split}
\label{eq:log_likelihood_orig}
\end{equation}
\begin{equation}
\begin{split}s_{m}^{\left(n\right)}\left(\mathbf{\boldsymbol{\Lambda}_{ori}}\right) & =\sum\limits _{k=1}^{K}\alpha_{k}e^{j\beta_{k}}e^{-j2\pi\left(f_{c,m}+nf_{s,m}\right)\left(\tau_{k}+\delta_{m}\right)}e^{j\phi_{m}},\end{split}
\label{eq:reconstruc_sig}
\end{equation}
where $s_{m}^{\left(n\right)}\left(\mathbf{\boldsymbol{\Lambda}_{ori}}\right)$
is the received signal reconstructed from the parameter $\mathbf{\boldsymbol{\Lambda}_{ori}}$.

In the next section, we shall propose a general Bayesian algorithm
to obtain the efficient approximation of marginal posteriori for the
target parameters.
\end{example}

\section{Parallel Stochastic Particle Variational Bayesian Inference for Non-Convex
Parameter Estimation}

\subsection{Problem Formulation for PSPVBI}

Since Problem $\mathcal{P}_{1}$ is, in general, non-convex, we focus
on designing an efficient algorithm to find a stationary point of
$\mathcal{P}_{1}$. There are two major challenges in solving Problem
$\mathcal{P}_{1}$: 1) there are generally many 'bad' local optimums
in the objective function due to the non-convexity and high-dimension
of the associated likelihood function; 2) it is difficult to obtain
a marginal posteriori distribution of the target parameter, due to
complex calculations of multiple integrals and tricky unstructured
distribution assumptions.

We propose a parallel stochastic particle variational Bayesian inference
(PSPVBI) algorithm to solve Problem $\mathcal{P}_{1}$, where variational
posteriori probability $q\left(\mathbf{\mathbf{\boldsymbol{\theta}}}_{j}\right)$
is approximated \cite{VBI+IS,PMD} by a variational particle set $\left\{ p_{j,n},w_{j,n}|\forall n=1,\ldots,N_{p}\right\} $
as follows
\begin{equation}
q\left(\mathbf{\mathbf{\boldsymbol{\theta}}}_{j};\mathbf{p}_{j},\mathbf{w}_{j}\right){\rm =}\sum\limits _{n=1}^{N_{p}}w_{j,n}\delta\left(\mathbf{\mathbf{\boldsymbol{\theta}}}_{j}-p_{j,n}\right),\label{eq:weighted sum-1}
\end{equation}
where $N_{p}$ denotes the number of particles. $\mathbf{p}_{j}=\left[p_{j,1},p_{j,2},...,p_{j,N_{p}}\right]^{T}$
and $\mathbf{w}_{j}=\left[w_{j,1},w_{j,2},...,w_{j,N_{p}}\right]^{T}$
are the positions (i.e. supporting point) and weights of the particles,
respectively. Particle positions can be initially drawn from a prior
$p_{j,n}^{\left(0\right)}\backsim p\left(\mathbf{\mathbf{\boldsymbol{\theta}}}_{j}\right),\forall j,n$,
and the initial weights can be set to be uniform, $w_{j,n}^{\left(0\right)}=\frac{1}{N_{p}},\forall j,n$.
Compared with structured continuous distribution assumptions (e.g.
Gaussian distribution, Gamma distribution, etc.), discrete particle
sets consisting of positions and weights can efficiently describe
the statistical probabilities of various variables within a given
region. In addition, they can also provide globally optimal posteriori
estimates and their associated belief compared to point estimates.

Given the particle approximation, we reformulate $\mathcal{P}_{1}$
to the following optimization problem $\mathcal{P}_{2}$:
\[
\begin{split}\mathcal{P}_{2}:\mathop{\min}\limits _{\mathbf{p},\mathbf{w}}\quad & \boldsymbol{L}\left(\mathbf{p},\mathbf{w}\right)\triangleq\int q\left(\mathbf{\mathbf{\boldsymbol{\theta}}};\mathbf{p},\mathbf{w}\right)ln\frac{q\left(\mathbf{\mathbf{\boldsymbol{\theta}}};\mathbf{p},\mathbf{w}\right)}{p\left(\boldsymbol{r}|\mathbf{\mathbf{\boldsymbol{\theta}}}\right)p\left(\mathbf{\mathbf{\boldsymbol{\theta}}}\right)}d\mathbf{\mathbf{\boldsymbol{\theta}}}\\
\text{s.t.}\quad & \sum\limits _{n=1}^{N_{p}}w_{j,n}=1,\quad\epsilon\leq w_{j,n}\leq1,\quad\forall j,n,\\
 & \hat{\mathbf{\mathbf{\boldsymbol{\theta}}}}_{j}-\Delta\hat{\mathbf{\mathbf{\boldsymbol{\theta}}}}_{j}/2\le p_{j,n}\le\hat{\mathbf{\mathbf{\boldsymbol{\theta}}}}_{j}+\Delta\hat{\mathbf{\mathbf{\boldsymbol{\theta}}}}_{j}/2,\quad\forall j,n,
\end{split}
\]
where $\epsilon>0$ is a small number. $\hat{\mathbf{\mathbf{\boldsymbol{\theta}}}}_{j}$
is the mean of the prior $p\left(\mathbf{\mathbf{\boldsymbol{\theta}}}_{j}\right)$
and $\Delta\hat{\mathbf{\mathbf{\boldsymbol{\theta}}}}_{j}$ is the
range determined by the prior $p\left(\mathbf{\mathbf{\boldsymbol{\theta}}}_{j}\right)$.
The truth value is highly likely to be located in the prior interval
$\left[\hat{\mathbf{\mathbf{\boldsymbol{\theta}}}}_{j}-\Delta\hat{\mathbf{\mathbf{\boldsymbol{\theta}}}}_{j}/2,\hat{\mathbf{\mathbf{\boldsymbol{\theta}}}}_{j}+\Delta\hat{\mathbf{\mathbf{\boldsymbol{\theta}}}}_{j}/2\right]$,
so the particle position is searched wherein to accelerate the convergence
rate. Note that it does not make sense to generate particles with
very small probabilities in approximate posteriori since these particles
contribute very little to the estimator. Therefore, we restrict the
probability of each particle $w_{j,n}$ to be larger than a small
number $\epsilon$.

In conventional PVBI, the position of particles are not updated. Besides,
a large number of particles are required to overcome the instability
caused by initial random sampling, and ensure that the estimation
locally converges to a 'good' stationary point. As such, the complexity
will rocket as the number of variables and particles increases.

Motivated by this observcation, we optimize the particle position
as well and further design an PSPVBI algorithm to solve the new problem
with much lower per-iteration complexity than the conventional PVBI
algorithm. Adding the optimization of particle positions $\mathbf{p}$
can improve the effectiveness of characterizing the target distribution
by discrete particles, and avoid the estimation result falling into
the local optimum due to poor initial sampling. Furthermore, updating
particle position can reduce the number of required particles, and
thus effectively reduce the computational overhead, which will be
discussed in detail in the following sections.

\subsection{PSPVBI Algorithm Design based on Parrallel SSCA\label{subsec:SPVBI-Algorithm-Design}}

Although updating particle positions can speed up convergence, multiple
integration in the objective function $\boldsymbol{L}\left(\mathbf{p},\mathbf{w}\right)$
of $\mathcal{P}_{2}$ is still intractable. To solve this problem,
we introduce the parallel stochastic successive convex approximations
(PSSCA) \cite{SSCA,PSSCA} to deal with the tricky crux of Bayesian
inference regarding multiple integrals.

According to the mean field theory \cite{meanfield}, the approximate
posteriori probability of each variable can be assumed to be independent
of each other, i.e. $q\left(\mathbf{\mathbf{\boldsymbol{\theta}}};\mathbf{p},\mathbf{w}\right)=\prod_{j=1}^{J}q\left(\mathbf{\boldsymbol{\theta}}_{j};\mathbf{p}_{j},\mathbf{w}_{j}\right)$,
so we have
\begin{align}
\int q\left(\mathbf{\mathbf{\boldsymbol{\theta}}}\right)lnq\left(\mathbf{\mathbf{\boldsymbol{\theta}}}\right)d\mathbf{\mathbf{\boldsymbol{\theta}}} & =\sum\limits _{j=1}^{J}\int q\left(\mathbf{\mathbf{\boldsymbol{\theta}}}_{j}\right)lnq\left(\mathbf{\boldsymbol{\theta}}_{j}\right)d\mathbf{\boldsymbol{\theta}}_{j}\nonumber \\
=\sum\limits _{j=1}^{J}\sum\limits _{n_{j}=1}^{N_{p}} & w_{j,n_{j}}lnw_{j,n_{j}}\triangleq C\left(\mathbf{w}\right).
\end{align}
We shall use $q\left(\mathbf{\mathbf{\boldsymbol{\theta}}}\right)$
as a simplified notation for $q\left(\mathbf{\mathbf{\boldsymbol{\theta}}};\mathbf{p},\mathbf{w}\right)$.
Further for the objective function:
\begin{align}
 & \boldsymbol{L}\left(\mathbf{p},\mathbf{w}\right)\nonumber \\
= & \sum\limits _{j=1}^{J}\int q\left(\mathbf{\mathbf{\boldsymbol{\theta}}}_{j}\right)lnq\left(\mathbf{\boldsymbol{\theta}}_{j}\right)d\mathbf{\boldsymbol{\theta}}_{j}-\int q\left(\mathbf{\mathbf{\boldsymbol{\theta}}}\right)\left[lnp\left(\boldsymbol{r}|\mathbf{\mathbf{\boldsymbol{\theta}}}\right)p\left(\mathbf{\mathbf{\boldsymbol{\theta}}}\right)\right]d\mathbf{\boldsymbol{\theta}}\nonumber \\
= & C\left(\mathbf{w}\right)-\left.\sum\limits _{n_{1}=1}^{N_{p}}\cdots\sum\limits _{n_{J}=1}^{N_{p}}\widetilde{w}\left[lnp\left(\boldsymbol{r}|\mathbf{\mathbf{\boldsymbol{\theta}}}\right)+lnp\left(\mathbf{\mathbf{\boldsymbol{\theta}}}\right)\right]\right|_{\mathbf{\mathbf{\boldsymbol{\theta}}}=\widetilde{\mathbf{p}}},\label{eq:L_expand}
\end{align}
where $\widetilde{w}=\prod_{j=1}^{J}w_{j,n_{j}}$ and $\widetilde{\mathbf{p}}=\left[p_{1,n_{1}},...p_{j,n_{j}}...,p_{J,n_{J}}\right]^{T}$.
As you can see, the second part of \eqref{eq:L_expand} involves a
summation of $N_{p}^{J}$ terms. Therefore, it is unacceptable to
directly solve $\mathcal{P}_{2}$ due to its exponential complexity.

To overcome this challenge, we rewrite the objective function as
\[
\boldsymbol{L}\left(\mathbf{p},\mathbf{w}\right)={\rm \mathbb{E}}_{q\left(\mathbf{\mathbf{\boldsymbol{\theta}}}\right)}\left[g\left(\mathbf{p},\mathbf{w};\mathbf{\mathbf{\boldsymbol{\theta}}}\right)\right],
\]
where ${\rm \mathbb{E}}_{q\left(\boldsymbol{\theta}\right)}\left[\cdot\right]$
represents the expectation operator over the variational distribution
$q\left(\mathbf{\mathbf{\boldsymbol{\theta}}}\right)$, and $g\left(\mathbf{p},\mathbf{w};\mathbf{\mathbf{\boldsymbol{\theta}}}\right){\rm =}C\left(\mathbf{w}\right)-\left[lnp\left(\boldsymbol{r}|\mathbf{\mathbf{\boldsymbol{\theta}}}\right)+lnp\left(\mathbf{\mathbf{\boldsymbol{\theta}}}\right)\right]$.
Therefore, $\mathcal{P}_{2}$ can be viewed as a stochastic optoimization
problem that involves a decision-dependent distribution, meaning that
the random state $\mathbf{\mathbf{\boldsymbol{\theta}}}$ follows
the distribution $q\left(\mathbf{\mathbf{\boldsymbol{\theta}}};\mathbf{p},\mathbf{w}\right)$,
which depends on the value of the optimization variables $\mathbf{p},\mathbf{w}$.
We shall use $q^{\left(t\right)}\left(\mathbf{\mathbf{\boldsymbol{\theta}}}\right)$
as a simplified notation for $q\left(\mathbf{\mathbf{\boldsymbol{\theta}}};\mathbf{p}^{\left(t\right)},\mathbf{w}^{\left(t\right)}\right)$
in the $t$-th iteration.

For SSCA applied in the stochastic optimization \cite{SSCA}, the
basic idea behind it is to iteratively optimize a sequence of convex
objective function (i.e. surrogate function) to approximate the original
non-convex one. To update the particle position $\mathbf{p}_{j}$
and particle weight $\mathbf{w}_{j}$ of the $j$-th variable $\mathbf{\boldsymbol{\theta}}_{j}$,
we can choose quadratic convex functions \cite{PSSCA,SSCA} as the
surrogate objective functions
\begin{equation}
\overline{f}_{p_{j}}^{\left(t\right)}\left(\mathbf{p}_{j}\right)=\left(\mathbf{\boldsymbol{{\rm f}}}_{p_{j}}^{\left(t\right)}\right)^{T}\left(\mathbf{p}_{j}-\mathbf{p}_{j}^{\left(t\right)}\right)+\varGamma_{p_{j}}^{\left(t\right)}\left\Vert \mathbf{p}_{j}-\mathbf{p}_{j}^{\left(t\right)}\right\Vert ^{2},\label{eq:surrFunc_p}
\end{equation}
\begin{equation}
\overline{f}_{w_{j}}^{\left(t\right)}\left(\mathbf{w}_{j}\right)=\left(\mathbf{\boldsymbol{{\rm f}}}_{w_{j}}^{\left(t\right)}\right)^{T}\left(\mathbf{w}_{j}-\mathbf{w}_{j}^{\left(t\right)}\right)+\Gamma_{w_{j}}^{\left(t\right)}\left\Vert \mathbf{w}_{j}-\mathbf{w}_{j}^{\left(t\right)}\right\Vert ^{2},\label{eq:surrFunc_w}
\end{equation}
where $\mathbf{\boldsymbol{{\rm f}}}_{p_{j}}^{\left(t\right)}$ and
$\mathbf{\boldsymbol{{\rm f}}}_{w_{j}}^{\left(t\right)}$ are unbiased
estimators of the gradient $\nabla_{\mathbf{p}_{j}}\boldsymbol{L}\left(\mathbf{p}^{\left(t\right)},\mathbf{w}^{\left(t\right)}\right)$
and $\nabla_{\mathbf{w}_{j}}\boldsymbol{L}\left(\mathbf{p}^{\left(t\right)},\mathbf{w}^{\left(t\right)}\right)$,
$\varGamma_{p_{j}}^{\left(t\right)}$ and $\Gamma_{w_{j}}^{\left(t\right)}$
are appropriately chosen positive stepsize. These surrogate subproblems
have closed-form solutions, which can be interpreted as the outcome
of gradient projection \cite{PSSCA,SPVBI}. Therefore, we can obtain
intermediate variables $\overline{\mathbf{p}}_{j}^{\left(t\right)},\overline{\mathbf{w}}_{j}^{\left(t\right)}$
by gradient projection as
\begin{equation}
\overline{\mathbf{p}}_{j}^{\left(t\right)}=Proj_{\Delta p}\left(\mathbf{p}_{j}^{\left(t\right)}-\Gamma_{p_{j}}^{\left(t\right)}\mathbf{\boldsymbol{{\rm f}}}_{p_{j}}^{\left(t\right)}\right),\label{eq:proj_p}
\end{equation}
\begin{equation}
\overline{\mathbf{w}}_{j}^{\left(t\right)}=Proj_{\Delta w}\left(\mathbf{w}_{j}^{\left(t\right)}-\Gamma_{w_{j}}^{\left(t\right)}\mathbf{\boldsymbol{{\rm f}}}_{w_{j}}^{\left(t\right)}\right),\label{eq:proj_w}
\end{equation}
where $Proj_{\Delta p/\Delta w}\left(\cdot\right)$ denotes the projection
operator related to the constraints in $\mathcal{P}_{2}$, which constrains
particle positions and weights within a priori interval or a simplex
and will be discussed in Section \ref{subsec:Module-design}. Finally,
the updated $\left\{ \mathbf{p}_{j}^{\left(t+1\right)},\mathbf{w}_{j}^{\left(t+1\right)}\right\} ,\forall j$
are given by
\begin{equation}
\mathbf{p}_{j}^{\left(t+1\right)}=\left(1-\gamma^{\left(t\right)}\right)\mathbf{p}_{j}^{\left(t\right)}+\gamma^{\left(t\right)}\overline{\mathbf{p}}_{j}^{\left(t\right)},\label{eq:x_update}
\end{equation}
\begin{equation}
\mathbf{w}_{j}^{\left(t+1\right)}=\left(1-\gamma^{\left(t\right)}\right)\mathbf{w}_{j}^{\left(t\right)}+\gamma^{\left(t\right)}\overline{\mathbf{w}}_{j}^{\left(t\right)},\label{eq:y_update}
\end{equation}
where $\gamma^{\left(t\right)}$ is a decreasing step size that will
be discussed later.

Next, we will give the construction of the unbiased estimators (i.e.
$\mathbf{\boldsymbol{{\rm f}}}_{p_{j}}^{\left(t\right)}$ and $\mathbf{\boldsymbol{{\rm f}}}_{w_{j}}^{\left(t\right)}$)
for the gradient of the objective function. The gradient of the objective
function can be converted to the following form:
\begin{align}
\nabla_{\mathbf{p}_{j}}\boldsymbol{L}\left(\mathbf{p}^{\left(t\right)},\mathbf{w}^{\left(t\right)}\right) & ={\rm \mathbb{E}}_{q^{\left(t\right)}\left(\mathbf{\mathbf{\boldsymbol{\theta}}}_{\sim j}\right)}\left[\nabla_{\mathbf{p}_{j}}g_{j}\left(\mathbf{p}_{j}^{\left(t\right)},\mathbf{w}_{j}^{\left(t\right)};\mathbf{\boldsymbol{\theta}}_{\sim j}\right)\right],\label{eq:grad_L_p}
\end{align}
\begin{align}
\nabla_{\mathbf{w}_{j}}\boldsymbol{L}\left(\mathbf{p}^{\left(t\right)},\mathbf{w}^{\left(t\right)}\right) & ={\rm \mathbb{E}}_{q^{\left(t\right)}\left(\mathbf{\mathbf{\boldsymbol{\theta}}}_{\sim j}\right)}\left[\nabla_{\mathbf{w}_{j}}g_{j}\left(\mathbf{p}_{j}^{\left(t\right)},\mathbf{w}_{j}^{\left(t\right)};\mathbf{\boldsymbol{\theta}}_{\sim j}\right)\right]\label{eq:grad_L_w}
\end{align}
where
\begin{align}
g_{j} & \left(\mathbf{p}_{j},\mathbf{w}_{j};\mathbf{\boldsymbol{\theta}}_{\sim j}\right){\rm =}C\left(\mathbf{w}\right)-\sum\limits _{n_{j}=1}^{N_{p}}w_{j,n_{j}}\nonumber \\
\times & \left[\ln p\left(p_{j,n_{j}}\right)+\ln p\left(\boldsymbol{r}\left|\mathbf{\boldsymbol{\theta}}_{\sim j},p_{j,n_{j}}\right.\right)\right],\label{eq:gj}
\end{align}
and $\mathbf{\boldsymbol{\theta}}_{\sim j}$ represents all the other
variables except $\mathbf{\boldsymbol{\theta}}_{j}$.

To approximate the gradients in (\ref{eq:grad_L_p}) and (\ref{eq:grad_L_w}),
in each iteration, based on the multiple samples of system state $\boldsymbol{\theta}$
generated by the distribution $q^{\left(t\right)}\left(\mathbf{\mathbf{\boldsymbol{\theta}}}\right)$,
$\mathbf{\boldsymbol{{\rm f}}}_{p_{j}}^{\left(t\right)}$ and $\mathbf{\boldsymbol{{\rm f}}}_{w_{j}}^{\left(t\right)}$
can be updated recursively as follows:
\begin{equation}
\begin{split}\mathbf{\boldsymbol{{\rm f}}}_{p_{j}}^{\left(t\right)} & =\left(1-\rho^{\left(t\right)}\right)\mathbf{\boldsymbol{{\rm f}}}_{p_{j}}^{\left(t-1\right)}+\frac{\rho^{\left(t\right)}}{B}\sum_{b=1}^{B}\nabla_{\mathbf{p}_{j}}g_{j}^{\left(t,b\right)},\end{split}
\label{eq:fx}
\end{equation}
\begin{equation}
\begin{split}\mathbf{\boldsymbol{{\rm f}}}_{w_{j}}^{\left(t\right)} & =\left(1-\rho^{\left(t\right)}\right)\mathbf{\boldsymbol{{\rm f}}}_{w_{j}}^{\left(t-1\right)}+\frac{\rho^{\left(t\right)}}{B}\sum_{b=1}^{B}\nabla_{\mathbf{w}_{j}}g_{j}^{\left(t,b\right)}\end{split}
,\label{eq:fy}
\end{equation}
where
\begin{align}
\nabla_{\mathbf{p}_{j}} & g_{j}^{\left(t,b\right)}=\left.\nabla_{\mathbf{p}_{j}}g_{j}\left(\mathbf{p}_{j},\mathbf{w}_{j}^{\left(t\right)};\mathbf{\boldsymbol{\theta}}_{\sim j}^{\left(b\right)}\right)\right|_{\mathbf{p}_{j}=\mathbf{p}_{j}^{\left(t\right)}}\nonumber \\
=vec & \left\{ -w_{j,n}^{\left(t\right)}\cdot\nabla_{p_{j,n}}\left[lnp\left(p_{j,n}^{\left(t\right)}\right)+lnp\left(\boldsymbol{r}\left|p_{j,n}^{\left(t\right)},\mathbf{\boldsymbol{\theta}}_{\sim j}^{\left(b\right)}\right.\right)\right]\right\} _{n=1}^{N_{p}},\label{eq:gx}
\end{align}
\begin{align}
\nabla_{\mathbf{w}_{j}} & g_{j}^{\left(t,b\right)}=\left.\nabla_{\mathbf{w}_{j}}g_{j}^{\left(t\right)}\left(\mathbf{p}_{j}^{\left(t\right)},\mathbf{w}_{j};\mathbf{\boldsymbol{\theta}}_{\sim j}^{\left(b\right)}\right)\right|_{\mathbf{w}_{j}=\mathbf{w}_{j}^{\left(t\right)}}\nonumber \\
=vec & \left[lnw_{j,n}^{\left(t\right)}+1-lnp\left(p_{j,n}^{\left(t\right)}\right)-lnp\left(\boldsymbol{r}\left|p_{j,n}^{\left(t\right)},\mathbf{\boldsymbol{\theta}}_{\sim j}^{\left(b\right)}\right.\right)\right]_{n=1}^{N_{p}}.\label{eq:gy}
\end{align}
$\left\{ \mathbf{\boldsymbol{\theta}}^{\left(b\right)}=\left[\mathbf{\boldsymbol{\theta}}_{1}^{\left(b\right)},\mathbf{\boldsymbol{\theta}}_{2}^{\left(b\right)},...,\mathbf{\boldsymbol{\theta}}_{J}^{\left(b\right)}\right]^{T},b=1,...,B\right\} $
is a mini-batch of $B$ samples generated by the distribution $\prod_{j=1}^{J}q\left(\mathbf{\boldsymbol{\theta}}_{j};\mathbf{p}_{j}^{\left(t\right)},\mathbf{w}_{j}^{\left(t\right)}\right)$,
and $\mathbf{w}^{\left(b\right)}$ is the corresponding particle weights
for the samples $\mathbf{\boldsymbol{\theta}}^{\left(b\right)}$.
$\rho^{\left(t\right)}$ is another decreasing step size that will
be discussed below and we set $\mathbf{\boldsymbol{{\rm f}}}_{p_{j}}^{\left(-1\right)}=0$,
$\mathbf{\boldsymbol{{\rm f}}}_{w_{j}}^{\left(-1\right)}=0$. It is
worth mentioning that both gradients and iterates (optimization variables)
are averaged over iteration. As such, under some technical conditions,
almost sure convergence to stationary points can be established.

To ensure the convergence of the algorithm, the step sizes $\rho^{\left(t\right)}$
and $\gamma^{\left(t\right)}$ must satisfy the following step-size
rules: 1) $\rho^{\left(t\right)}\rightarrow0$, $\sum_{t}\rho^{\left(t\right)}=\infty$,
$\sum_{t}\left(\rho^{\left(t\right)}\right)^{2}<\infty$; 2) $\underset{t\rightarrow\infty}{\lim}\gamma^{\left(t\right)}/\rho^{\left(t\right)}=0$.
A typical choice of $\rho^{\left(t\right)},\gamma^{\left(t\right)}$
is $\rho^{\left(t\right)}=\mathcal{O}\left(t^{-\kappa_{1}}\right)$,
$\gamma^{\left(t\right)}=\mathcal{O}\left(t^{-\kappa_{2}}\right)$,
where $0.5<\kappa_{1}<\kappa_{2}\leq1$. Such form of step sizes have
been widely considered in stochastic optimization \cite{PSSCA}. The
proposed PSPVBI is guaranteed to converge to stationary points of
the Problem $\mathcal{P}_{2}$, as will be proved in Section \ref{sec:Convergence-Analysis}.
After the convergence, the corresponding discrete distribution of
each variable $q\left(\mathbf{\boldsymbol{\theta}}_{j}\right)$ composed
of particles will approximate the marginal posteriori distribution.
As a result, we can take the particle position with the highest probability
or the weighted sum of the particles as the final estimate, which
are the approximate MAP estimate and MMSE estimate, respectively.

\subsection{Summary of the PSPVBI Algorithm}

The overall PSPVBI algorithm is summarized in Algorithm \ref{alg:SPVBI},
with its improvements mainly in three aspects. 1) \textbf{Flexibility:}
Unlike other restricted parametric variational inference algorithms,
PSPVBI approximates the posteriori in a non-parametric form (i.e.,
discrete particles). Higher flexibility in density space makes PSPVBI
less prone to getting trapped in local optima, thereby leading to
favorable convergence results. Additionally, the update of particle
positions provides the algorithm with greater degrees of freedom,
enabling it to further enhance performance, accelerate convergence,
and reduce the number of particles required. 2) \textbf{Efficiency:}
To avoid the exponential complexity of multiple integrals in high-dimensional
VBI, we apply the PSSCA approach in \cite{SSCA,PSSCA} to improve
the sampling efficiency using the average-over-iteration technique.
This trick of mini-batch sampling and smoothed recursion avoids scanning
over the whole probability space in each update. Given the aforementioned
flexibility and efficiency of PSPVBI, we are more likely to obtain
accurate global posteriori estimates for high-dimensional, non-convex
parameter estimation problems. 3) \textbf{Generality:} We extend the
SPVBI in \cite{SPVBI} from multi-band sensing to more general parameter
estimation scenes, and from alternating optimization to parallel optimization
to facilitate deep unfolding and convergence acceleration.

As an extension of existing SSCA in \cite{SSCA}, with the update
of particles in each iteration, the distribution of random states
is no longer constant, but changes dynamically, i.e., is control dependent.
Nevertheless, under certain technical conditions, convergence can
still be guaranteed by constructing a series of parallel sub-surrogate
functions. Please refer to Section \ref{subsec:ConvProof_SPVBI} for
the proof.
\begin{algorithm}[tbh]
\caption{\label{alg:SPVBI}Parallel Stochastic Particle VBI Algorithm}

\textbf{Input}: Measurement data, prior $p\left(\mathbf{\widehat{\boldsymbol{\theta}}}\right)$,
$\left\{ \rho^{\left(t\right)},\gamma^{\left(t\right)}\right\} $.

\textbf{Initialization}: $\left\{ p_{j,n}^{\left(0\right)}\backsim p\left(\mathbf{\widehat{\boldsymbol{\theta}}}_{j}\right),w_{j,n}^{\left(0\right)}=\frac{1}{N_{p}}\right\} _{n=1}^{N_{p}},\forall j$.

\textbf{While not converge do} $(t\rightarrow$$\infty)$

$\quad$ \textbf{Parallel execution for all $J$ variables:}

$\quad\quad$Generate a mini-batch of realization based on the

$\quad\quad$$\left\{ p_{j,n}^{\left(t\right)},w_{j,n}^{\left(t\right)}\right\} _{n=1}^{N_{p}},\forall j$;

$\quad\quad$Update $\mathbf{\boldsymbol{{\rm f}}}_{p_{j}}^{\left(t\right)}$
and $\mathbf{\boldsymbol{{\rm f}}}_{w_{j}}^{\left(t\right)},\forall j$;

$\quad\quad$Solving surrogate optimization problems;

$\quad\quad$Update the particle position and weight of variable $\mathbf{\boldsymbol{\theta}}_{j},\forall j$;

\textbf{end}

\textbf{Output}: Find the position $p_{j}^{*},j=1,2,\ldots J$ of
the particle with the maximum weight $w_{j}^{\left(max\right)}$.
(i.e. MAP estimation)
\end{algorithm}

\section{Proposed Deep-Unfolding Network}

As a model-driven approach, the PSPVBI algorithm can leverage domain
knowledge relevant to the problem, such as signal models and prior
information, providing certain performance guarantees. However, it
faces challenges during the iterative process, including difficulties
in hyperparameter tuning and slow convergence. In contrast, general
deep neural networks (DNN) have the capability to automatically learn
network parameters through training, exhibiting powerful fitting capabilities
\cite{2D_ResFreq}. However, they suffer from limited interpretability
and generalization abilities. To combine the strengths of both approaches,
deep unfolding (DU) techniques in machine learning have been recently
introduced \cite{Survey_DU1,Survey_DU2}.

DU networks combine the expressive power of conventional DNNs with
the internal structure of model-based methods, while allowing for
inference to be performed within a fixed number of layers that can
be optimized for best performance. Specifically, DU networks unfold
model-driven algorithms with a fixed number of iterations $T$, into
a layer-wise structure analogous to a neural network with trainable
parameters. These parameters are then learned using deep learning
(DL) techniques (e.g. suitable loss functions, stochastic gradient
descent, and backpropagation). The resulting unfolded algorithm requires
fewer iterations and yields optimized network parameters, exhibiting
significant advantages. The module design corresponding to Learnable
PSPVBI (LPSPVBI) is shown below.

\subsection{Module design\label{subsec:Module-design}}

\begin{figure}[tp]
\begin{centering}
\textsf{\includegraphics[scale=0.23]{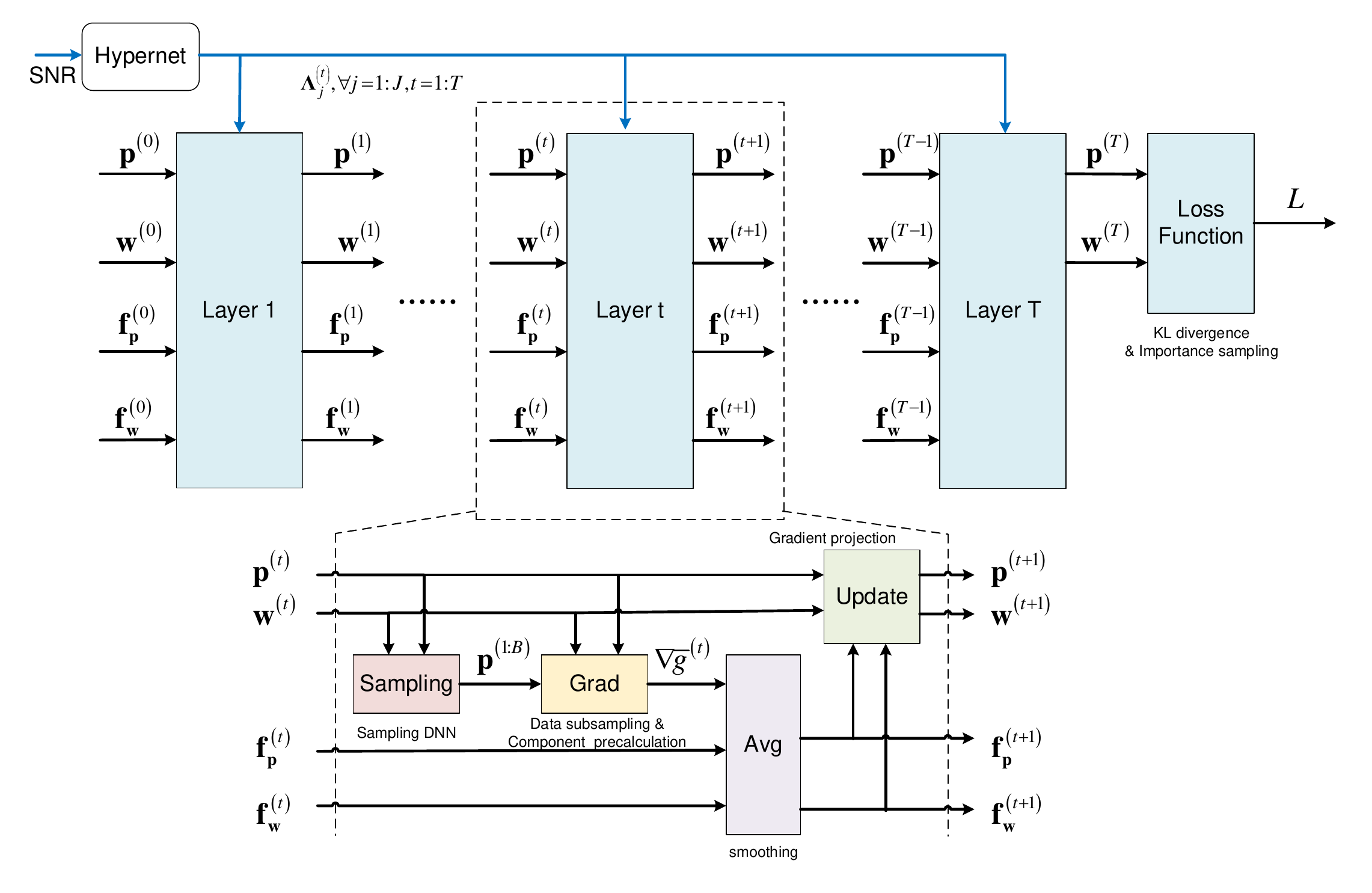}}
\par\end{centering}
\caption{\textsf{\label{Architecture}}The architecture of the deep-unfolding
framework.}
\end{figure}

The overall network structure is illustrated in Fig. \ref{Architecture},
consisting of $T$ layers cascaded together. Each layer corresponds
to one iteration of the original algorithm, including a complete update
process of variational distributions for all $J$ variables. Each
layer has the same structure but different trainable parameters\footnote{The smoothing effect of $\rho^{\left(t\right)}$ on gradients $\mathbf{\boldsymbol{{\rm f}}}_{p_{j}/w_{j}}^{\left(t\right)}$
can be partially achieved by tuning $\Gamma_{p_{j}/w_{j}}^{\left(t\right)}$
as a multiplier of $\mathbf{\boldsymbol{{\rm f}}}_{p_{j}/w_{j}}^{\left(t\right)}$
in \eqref{eq:proj_p}, \eqref{eq:proj_w}, and since $\Gamma_{p_{j}/w_{j}}^{\left(t\right)}$
can directly determine $\overline{\mathbf{p}}_{j}^{\left(t\right)}/\overline{\mathbf{w}}_{j}^{\left(t\right)}$,
the smoothing effect of $\gamma^{\left(t\right)}$ on variables can
also be partially achieved by tuning $\Gamma_{p_{j}/w_{j}}^{\left(t\right)}$.
For the purpose of reducing the hyperparameter search space, other
step sizes like $\left\{ \rho^{\left(t\right)},\gamma^{\left(t\right)}\right\} $
are not set as trainable parameters, but are fixed to a set of good
empirical values.} $\mathbf{\mathbf{\boldsymbol{\varLambda}}}_{j}^{\left(t\right)}=\left[\Gamma_{w_{j}}^{\left(t\right)},\Gamma_{p_{j}}^{\left(t\right)}\right]^{T},\forall j=1:J,t=1:T$.
Particle positions $\mathbf{p}$, weights $\mathbf{w}$, and gradients
for updating positions and weights $\boldsymbol{{\rm f}}_{\mathbf{p}},\boldsymbol{{\rm f}}_{\mathbf{w}}$
are passed from one layer to another, with each of them being an $\left(J\cdot N_{p}\right)$-dimensional
vector.

Within each layer, there are four specific modules: 1) sampling module,
2) gradient calculation module, 3) gradient smoothing module, and
4) particle update module. Each module will be introduced in detail
below.

\subsubsection{Sampling module}

When approximating the gradient in \eqref{eq:fx} and \eqref{eq:fy},
it is necessary to randomly sample in the state space according to
the current discrete variational distributions. This sampling operation
should be differentiable due to the backpropagation required by the
DU network. A common approach is to use the Gumbel-Softmax method
\cite{Gumbel_Softmax}, which replaces random nodes with deterministic
nodes plus random noise. To some extent, this approximation can mimic
random sampling. However, this method introduces significant randomness
when generating forward samples and computing backward gradients,
resulting in a large variance and making it difficult for the network
to converge. Even worse, the logarithmic operator in gradient computation
will exacerbate this issue, as the inputs to the logarithmic function
are probabilities ranging from $0$ to $1$. Therefore, it is necessary
to generate a sufficient number of samples to ensure that the sampled
statistical distribution approximates the probability distribution,
resulting in increased computational overhead for each layer.

\begin{figure}[htbp]
\begin{centering}
\textsf{\includegraphics[scale=0.3]{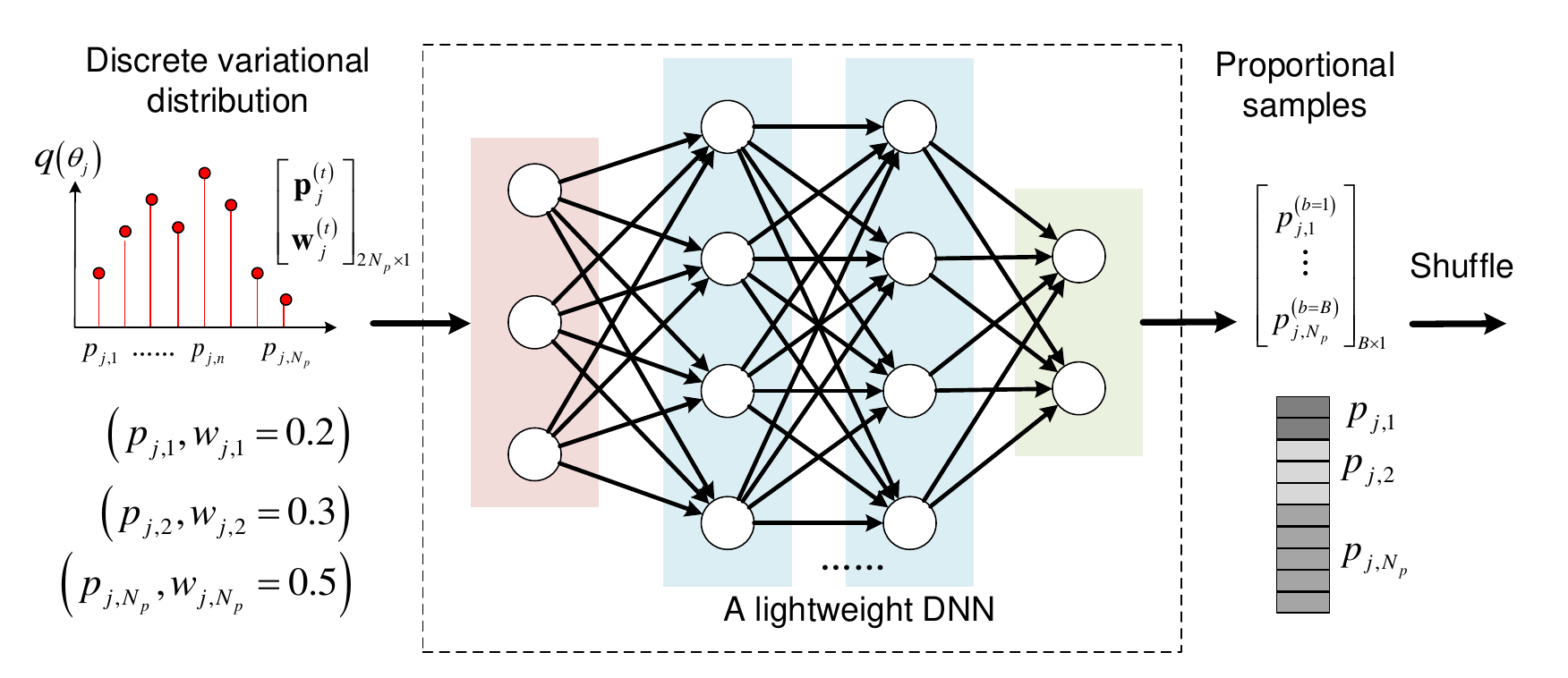}}
\par\end{centering}
\caption{\textsf{\label{DNN_sample}}Illustration of the sampling network.}
\end{figure}

Since the function mapping that generates a specific number of samples
based on the discrete distribution is deterministic, we consider using
a simple pre-trained lightweight DNN network as shown in Fig. \ref{DNN_sample}
to approximate this mapping. The input of DNN is a discrete distribution
$q\left(\mathbf{\boldsymbol{\theta}}_{j}\right)$, consisting of particle
positions and particle weights $\left\{ p_{j,n}^{\left(t\right)},w_{j,n}^{\left(t\right)}\right\} _{n=1}^{N_{p}}$,
and the output is $B$ proportional sample values. For example, set
$B=10$, if input $\left\{ \left(p_{j,1}^{\left(t\right)},0.2\right),\left(p_{j,2}^{\left(t\right)},0.3\right),\left(p_{j,3}^{\left(t\right)},0.5\right)\right\} $,
the output is $\left\{ \underset{2}{\underbrace{p_{j,1}^{\left(t\right)}\ldots}},\underset{3}{\underbrace{p_{j,2}^{\left(t\right)}\ldots}},\underset{5}{\underbrace{p_{j,3}^{\left(t\right)}\ldots}}\right\} $.
Then, the sampled values of different variables are shuffled to calculate
the gradient. As we all know, DNNs are capable of backpropagation,
which satisfies the requirements for subsequent hyperparameter training.
Additionally, proportional mapping can effectively indicate the current
probability distribution without the need for a large number of random
samples.

\subsubsection{Gradient calculation module}

Based on \eqref{eq:gx}, \eqref{eq:gy} and samples $\mathbf{\boldsymbol{\theta}}_{j}^{\left(b\right)},\forall j,b$,
we can obtain the average gradients of position and weight under several
realizations
\begin{equation}
\nabla_{\mathbf{p}_{j}}\overline{g}^{\left(t\right)}=\frac{1}{B}\stackrel[b=1]{B}{\sum}\nabla_{\mathbf{p}_{j}}g^{\left(t,b\right)},\label{eq:avg_p}
\end{equation}
\begin{equation}
\nabla_{\mathbf{w}_{j}}\overline{g}^{\left(t\right)}=\frac{1}{B}\stackrel[b=1]{B}{\sum}\nabla_{\mathbf{w}_{j}}g_{j}^{\left(t,b\right)}.\label{eq:avg_w}
\end{equation}
The major computation bottleneck in \eqref{eq:gx}, \eqref{eq:gy}
lies on calculating $\ln p\left(\boldsymbol{r}\left|\mathbf{\boldsymbol{\theta}}_{\sim j}^{\left(b\right)},p_{j,n}^{\left(t\right)}\right.\right)$
and $\nabla_{p_{j,n}}lnp\left(\boldsymbol{r}\left|\mathbf{\boldsymbol{\theta}}_{\sim j}^{\left(b\right)},p_{j,n}^{\left(t\right)}\right.\right)$.
Considering that most variables remain fixed as $\mathbf{\boldsymbol{\theta}}_{\sim j}^{\left(b\right)}$
for different particles $n=1,\ldots,N_{p}$, it is possible to precompute
certain components and store them for later use during traversal.
Furthermore, for high-dimensional $\boldsymbol{r}=\left[r_{1},r_{2},\ldots,r_{N}\right]^{T}$,
we can approximate the likelihood function $lnp\left(\boldsymbol{r}|\mathbf{\mathbf{\boldsymbol{\theta}}}\right)$
with subsampled minibatchs $\Omega\subset\left\{ 1,\ldots,N\right\} $
of the data \cite{SteinVI}
\begin{equation}
lnp\left(\boldsymbol{r}|\mathbf{\mathbf{\boldsymbol{\theta}}}\right)\approx\frac{N}{\left|\Omega\right|}\underset{n\in\Omega}{\sum}lnp\left(\boldsymbol{r}_{n}|\mathbf{\mathbf{\boldsymbol{\theta}}}\right).
\end{equation}
As a result, a significant amount of redundant and computationally
expensive calculations can be avoided, making the computational cost
affordable even in big data settings.

\subsubsection{Gradient smoothing module}

The gradient smoothing module is employed to adjust the weighting
between historical gradients and the current gradients as shown in
\eqref{eq:fx} and \eqref{eq:fy}. As the iterations advance, the
smoothed gradient effectively approximates the true gradient, even
when only a small number of samples are used for gradient averaging.

\subsubsection{Particle update module}

For updating particle positions,
\begin{equation}
\mathbf{p}_{j}^{\left(t+1\right)}=\left(1-\gamma^{\left(t\right)}\right)\mathbf{p}_{j}^{\left(t\right)}+\gamma^{\left(t\right)}Proj_{\Delta p}\left(\mathbf{p}_{j}^{\left(t\right)}-\Gamma_{p_{j}}^{\left(t\right)}\mathbf{\boldsymbol{{\rm f}}}_{p_{j}}^{\left(t\right)}\right).\label{eq:x_update-2}
\end{equation}
Due to the quadratic surrogate function and box constraints in $\mathcal{P}_{2}$,
the projection operation $Proj_{\Delta p}\left(\cdot\right)$ can
be expressed as a simple piecewise function
\begin{equation}
Proj_{\Delta p}\left(x\right)=\begin{cases}
a & x\in\left(-\infty,a\right)\\
x & x\in\left[a,b\right]\\
b & x\in\left(b,+\infty\right)
\end{cases},
\end{equation}
where $a=\hat{\mathbf{\mathbf{\boldsymbol{\theta}}}}_{j}-\Delta\hat{\mathbf{\mathbf{\boldsymbol{\theta}}}}_{j}/2$,
$b=\hat{\mathbf{\mathbf{\boldsymbol{\theta}}}}_{j}+\Delta\hat{\mathbf{\mathbf{\boldsymbol{\theta}}}}_{j}/2$.
Its derivative is $\nabla Proj_{\Delta p}\left(x\right)=\begin{cases}
1 & x\in\left[a,b\right]\\
0 & otherwise
\end{cases}$.

For updating particle weights,
\begin{equation}
\mathbf{w}_{j}^{\left(t+1\right)}=\left(1-\gamma^{\left(t\right)}\right)\mathbf{w}_{j}^{\left(t\right)}+\gamma^{\left(t\right)}Proj_{\Delta w}\left(\mathbf{w}_{j}^{\left(t\right)}-\Gamma_{w_{j}}^{\left(t\right)}\mathbf{\boldsymbol{{\rm f}}}_{w_{j}}^{\left(t\right)}\right).\label{eq:y_update-2}
\end{equation}
Since the constraint conditions of the subproblem are coupled simplex
constraints in $\mathcal{P}_{2}$, the projection operation can be
iteratively accomplished using the intuitive alternating projection
method. It should be noted that the alternating projection method
is not the most efficient method for solving this problem \cite{ProjSimplex}.
However, considering that the entire projection process needs to support
gradient backpropagation, it was suitbale to use this approach in
the DU network. We define sets $\mathbb{C}_{1}=\left\{ \mathbf{w}_{j}\in\mathbb{R}^{N_{p}}:\mathbf{w}_{j}\cdot\mathbf{1}_{N_{p}}^{T}=1\right\} $
and $\mathbb{C}_{2}=\left\{ \mathbf{w}_{j}\in\mathbb{R}^{N_{p}}:w_{j,n}\geq\epsilon,n=1,\ldots,N_{p}\right\} $,
where $\mathbf{1}_{N_{p}}$ is a $N_{p}$-dimensional vector of ones.
By alternately projecting onto set $\mathbb{C}_{1}$ and set $\mathbb{C}_{2}$,
after a few steps, we can obtain the result projected onto the simplex
\cite{Alt_proj}. Details of the alternating projection are provided
below.
\begin{algorithm}[tbh]
\caption{\label{alg:Alter_Proj}Alternating Projection Method.}

\textbf{Input}: $\mathbf{w}_{j}$.\textbf{ }(Initialization: $\mathbf{y}_{r}=\mathbf{y}_{r+1}=\mathbf{w}_{j},error=+\infty,\epsilon\rightarrow0$)

\textbf{While }$\left(error>10^{-6}\right)$

$\quad\quad$$\mathbf{y}_{r+1}=\max\left\{ \left[\mathbf{y}_{r}-\frac{1}{N_{p}}\left(\mathbf{y}_{r}\cdot\mathbf{1}_{N_{p}}^{T}-1\right)\mathbf{1}_{N_{p}}\right],\epsilon\cdot\mathbf{1}_{N_{p}}\right\} $

$\quad\quad$$error=\left\Vert \mathbf{y}_{r+1}-\mathbf{y}_{r}\right\Vert _{2}$;

$\quad\quad$$\mathbf{y}_{r}=\mathbf{y}_{r+1}$;

\textbf{end}

\textbf{Output}: $\mathbf{y}_{r}$.
\end{algorithm}

Hence, its derivative is $\nabla Proj_{\Delta\mathbf{w}}\left(\mathbf{w}\right)=\left[\left(\mathbf{I}_{N_{p}\times N_{p}}-\frac{1}{N_{p}}\right)diag\left(I\left(\mathbf{w}\right)\right)\right]^{iter}$,
$I\left(x\right)=\begin{cases}
1 & x\geq\epsilon\\
0 & x<\epsilon
\end{cases}$, where $iter$ represents the number of iterations for alternating
projection, $\mathbf{I}_{N_{p}\times N_{p}}$ is a $N_{p}\times N_{p}$-dimensional
identity matrix.

\subsubsection{Hypernetwork design\label{subsec:Hypernetwork-design}}

Despite the LPSPVBI network achieving considerable estimation performance
with faster convergence after training, there is still performance
degradation due to a mismatch between the training and testing environments.
This occurs because the variables learned under specific conditions
are not suitable for different scenarios. For example, if the LPSPVBI
is trained using data generated with SNR = $15$dB, its performance
will decline when tested with SNR = $10$dB. To address this issue,
we introduce the Hypernetwork \cite{HyperNet} into the LPSPVBI framework.

A hypernetwork, short for 'hyperparameter network', is a type of neural
network architecture used in deep learning. It is designed to dynamically
generate or learn the parameters of another neural network (i.e. target
network) based on input data. Due to the high correlation between
the training parameters and the $SNR$ (or noise precision), we use
the $SNR$ and training parameters $\mathbf{\mathbf{\boldsymbol{\varLambda}}}\in\mathbb{R}^{2J\cdot T\times1}$
as the inputs and outputs of the hypernetwork. Subsequently, a classical
three-layer fully connected hypernetwork is adopted to learn such
a mapping in order to enhance the generalization capability of LPSPVBI
in different noise environments, which can be written as
\begin{equation}
\mathbf{\mathbf{\boldsymbol{\varLambda}}}=\mathbf{W}_{3}\cdot ReLU\left(\mathbf{W}_{2}\cdot ReLU\left(\mathbf{W}_{1}\cdot SNR+\mathbf{b}_{1}\right)+\mathbf{b}_{2}\right)+\mathbf{b}_{3},
\end{equation}
where $\mathbf{W}_{1,2,3}$ and $\mathbf{b}_{1,2,3}$ are the parameters
to be trained for the Hypernetwork.

In addition, for the parameter estimation scenario, the number of
variables $J$ is usually not constant, such as the number of paths
and the targets. To make the unfolded algorithm adaptive to different
$J$, we can train our network with a larger $J_{0}$ in advance,
so that it can be straightforwardly transferred to the scenarios with
$J\leq J_{0}$. In the inference stage, we only need to set $\left\{ \mathbf{p}_{j}^{\left(0\right)},\mathbf{w}_{j}^{\left(0\right)},\boldsymbol{{\rm f}}_{p,j}^{\left(0\right)},\boldsymbol{{\rm f}}_{w,j}^{\left(0\right)}\right\} _{j=J+1}^{J_{0}}=0$
when $J<J_{0}$.

\subsection{Training Details}

\subsubsection{Loss function}

As the objective function of the PSPVBI algorithm in $\mathcal{P}_{2}$,
the KL divergence is defined as the loss function of DU, which measures
the distance between the variational distribution $q\left(\mathbf{\mathbf{\boldsymbol{\theta}}}\right)$
after $T$ iterations and the true posteriori distribution $p\left(\mathbf{\mathbf{\boldsymbol{\theta}}}|\boldsymbol{r}\right)$:
\begin{align}
Loss & \triangleq\int q\left(\mathbf{\mathbf{\boldsymbol{\theta}}};\mathbf{p}^{\left(T\right)},\mathbf{w}^{\left(T\right)}\right)\ln\frac{q\left(\mathbf{\mathbf{\boldsymbol{\theta}}};\mathbf{p}^{\left(T\right)},\mathbf{w}^{\left(T\right)}\right)}{p\left(\boldsymbol{r}|\mathbf{\mathbf{\boldsymbol{\theta}}}\right)p\left(\mathbf{\mathbf{\boldsymbol{\theta}}}\right)}d\mathbf{\mathbf{\boldsymbol{\theta}}}.\label{eq:LossFunc}
\end{align}
Similar to the approach of approximating gradients through sampling,
we can adopt Monte Carlo sampling to approximate the above multiple
integral. To improve the accuracy of the loss function, it is advisable
to increase the number of samples used, for example, $B=500$.

\subsubsection{Back propagation}

According to the loss function in \eqref{eq:LossFunc}, we can first
obtain the gradients with respect to the final layer's particle positions
and weights:
\begin{align}
\frac{\partial Loss}{\partial\mathbf{p}_{j}^{\left(T\right)}} & \triangleq{\rm \mathbb{E}}_{q\left(\mathbf{\mathbf{\boldsymbol{\theta}}};\mathbf{p}^{\left(T\right)},\mathbf{w}^{\left(T\right)}\right)}\left[\frac{\partial g^{\left(T\right)}}{\partial\mathbf{p}_{j}^{\left(T\right)}}\right]\nonumber \\
\approx\frac{1}{B_{grad}} & \sum\limits _{b=1}^{B_{grad}}\nabla_{\mathbf{p}_{j}}g_{j}^{\left(T\right)}\left(\mathbf{p}_{j}^{\left(T\right)};\mathbf{w}_{j}^{\left(T\right)},\mathbf{\boldsymbol{\theta}}_{\sim j}^{\left(b\right)}\right),\label{eq:grad_T_p}
\end{align}
\begin{align}
\frac{\partial Loss}{\partial\mathbf{w}_{j}^{\left(T\right)}} & \triangleq{\rm \mathbb{E}}_{q\left(\mathbf{\mathbf{\boldsymbol{\theta}}};\mathbf{p}^{\left(T\right)},\mathbf{w}^{\left(T\right)}\right)}\left[\frac{\partial g^{\left(T\right)}}{\partial\mathbf{w}_{j}^{\left(T\right)}}\right]\nonumber \\
\approx\frac{1}{B_{grad}} & \sum\limits _{b=1}^{B_{grad}}\nabla_{\mathbf{w}_{j}}g_{j}^{\left(T\right)}\left(\mathbf{w}_{j}^{\left(T\right)};\mathbf{p}_{j}^{\left(T\right)},\mathbf{\boldsymbol{\theta}}_{\sim j}^{\left(b\right)}\right).\label{eq:grad_T_w}
\end{align}
After obtaining the final layer's gradients, we can perform gradient
backpropagation using the chain rule of differentiation to get other
layers' gradients $\frac{\partial\boldsymbol{L}oss}{\partial\mathbf{p}_{j}^{\left(t\right)}},\frac{\partial\boldsymbol{L}oss}{\partial\mathbf{w}_{j}^{\left(t\right)}},\frac{\partial\boldsymbol{L}oss}{\partial\mathbf{\boldsymbol{{\rm f}}}_{p_{j}}^{\left(t\right)}},\frac{\partial\boldsymbol{L}oss}{\partial\mathbf{\boldsymbol{{\rm f}}}_{w_{j}}^{\left(t\right)}},$$\forall t=\left(T-1\right),\ldots,2,1$.
Then, we can further obtain the gradients of the loss function with
respect to the hyperparameters $\mathbf{\mathbf{\boldsymbol{\varLambda}}}_{j}^{\left(t\right)}$
of each layer
\begin{equation}
\frac{\partial Loss}{\partial\mathbf{\mathbf{\boldsymbol{\varLambda}}}_{j}^{\left(t\right)}}=\frac{\partial Loss}{\partial\mathbf{p}_{j}^{\left(t+1\right)}}\cdot\frac{\partial\mathbf{p}_{j}^{\left(t+1\right)}}{\partial\mathbf{\mathbf{\boldsymbol{\varLambda}}}_{j}^{\left(t\right)}}+\frac{\partial Loss}{\partial\mathbf{w}_{j}^{\left(t+1\right)}}\cdot\frac{\partial\mathbf{w}_{j}^{\left(t+1\right)}}{\partial\mathbf{\mathbf{\boldsymbol{\varLambda}}}_{j}^{\left(t\right)}},\label{eq:grad_step}
\end{equation}
where $\frac{\partial\mathbf{p}_{j}^{\left(t+1\right)}}{\partial\Gamma_{p_{j}}^{\left(t\right)}}=\gamma^{\left(t\right)}\left(-\mathbf{\boldsymbol{{\rm f}}}_{w_{j}}^{\left(t\right)}\right)\nabla Proj_{\Delta w}$,
$\frac{\partial\mathbf{p}_{j}^{\left(t+1\right)}}{\partial\Gamma_{w_{j}}^{\left(t\right)}}=0$,
$\frac{\partial\mathbf{w}_{j}^{\left(t+1\right)}}{\partial\Gamma_{p_{j}}^{\left(t\right)}}=0$,
$\frac{\partial\mathbf{w}_{j}^{\left(t+1\right)}}{\partial\Gamma_{w_{j}}^{\left(t\right)}}=\gamma^{\left(t\right)}\left(-\mathbf{\boldsymbol{{\rm f}}}_{p_{j}}^{\left(t\right)}\right)\nabla Proj_{\Delta p}$.
Finally, we use the gradients as a loss function of the Hypernetwork
described in \ref{subsec:Hypernetwork-design} to train the parameters
$\mathbf{W}_{1,2,3}$ and $\mathbf{b}_{1,2,3}$.

By conducting practical tests or simulating different wireless environments
(i.e. SNR, delay path, etc.), we can obtain a large amount of training
data. Additionally, training techniques from machine learning, such
as mini-batch averaging, stochastic gradient descent (SGD), Adam optimization
\cite{Adam}, and others, can be applied to train the hyperparameters
of the DU network. The detailed training procedures of the LPSPVBI
are presented in Algorithm \ref{alg:LSPVBI}.
\begin{algorithm}[tbh]
\caption{\label{alg:LSPVBI}The training procedures of PSPVBI-Inspired Deep-Unfolding
Network.}

Given the training set. Set the number of layers $T$, the batch size
$N_{batch}$, the tolerance of accuracy, the maximum iteration number
$I_{max}$, and the current iteration index of the training stage
$iter=0$.

\textbf{Repeat}

$\quad\quad$\textbf{1. Forward propagation:} Select a group of samples
from the training set and initialize $\left\{ \mathbf{p}^{\left(0\right)},\mathbf{w}^{\left(0\right)},\boldsymbol{{\rm f}}_{\mathbf{p}}^{\left(0\right)},\boldsymbol{{\rm f}}_{\mathbf{w}}^{\left(0\right)}\right\} $.
Compute $\left\{ \mathbf{p}^{\left(t\right)},\mathbf{w}^{\left(t\right)},\boldsymbol{{\rm f}}_{\mathbf{p}}^{\left(t\right)},\boldsymbol{{\rm f}}_{\mathbf{w}}^{\left(t\right)}\right\} ,\forall t=1,\ldots,T$
based on \eqref{eq:proj_p}-\eqref{eq:y_update}.

$\quad\quad$\textbf{2. }Then plug $\left\{ \mathbf{p}^{\left(T\right)},\mathbf{w}^{\left(T\right)}\right\} $
into the loss function and obtain its value.

$\quad\quad$\textbf{3. Backward propagation:} Firstly, compute the
gradients with respect to variables $\left\{ \mathbf{p}^{\left(T\right)},\mathbf{w}^{\left(T\right)}\right\} $
in the last layer based on \eqref{eq:grad_T_p}-\eqref{eq:grad_T_w}.
Secondly, compute the gradients of $\left\{ \mathbf{p}^{\left(t\right)},\mathbf{w}^{\left(t\right)},\boldsymbol{{\rm f}}_{\mathbf{p}}^{\left(t\right)},\boldsymbol{{\rm f}}_{\mathbf{w}}^{\left(t\right)}\right\} ,$$\forall t=\left(T-1\right),\ldots,2,1$
according to the chain rule. Finally, compute the gradients of trainable
parameters and based on \eqref{eq:grad_step}.

$\quad\quad$\textbf{4. Update Hypernetwork parameters:} Repeat steps
$1-3$ for $N_{batch}$ times and compute the average gradients of
trainable parameters in a batch. Then, apply mini-batch SGD to update
the Hypernetwork parameters.

$\quad\quad$\textbf{5. }$iter=iter+1$.

\textbf{Until} The loss function in the validation set converges or
$iter\geq I_{max}$.
\end{algorithm}

\section{Convergence Analysis\label{sec:Convergence-Analysis}}

In this section, we will present the convergence analysis of the PSPVBI
algorithm and the deep-unfolded LPSPVBI algorithm.

\subsection{Convergence analysis of PSPVBI\label{subsec:ConvProof_SPVBI}}

In this subsection, we show that the proposed PSSCA-based PSPVBI algorithm
can be guaranteed to converge to a stationary solution of $\mathcal{P}_{2}$.

\begin{figure}[htbp]
\begin{centering}
\textsf{\includegraphics[scale=0.6]{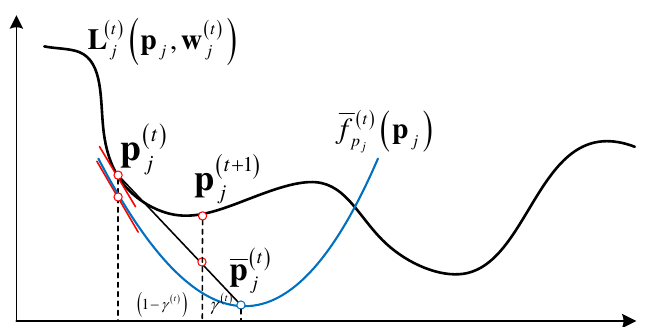}}
\par\end{centering}
\caption{\textsf{\label{SCA}}An illustration of stochastic successive convex
approximation.}
\end{figure}

Firstly, we point out that the gradient projection operation \eqref{eq:proj_p},
\eqref{eq:proj_w} in PSPVBI can be viewed as solving a constrained
quadratic surrogate subproblem, as illustrated in Fig. \ref{SCA}.
Moreover, the convex quadratic surrogate functions $\overline{f}_{p_{j}}^{\left(t\right)}\left(\mathbf{p}_{j}\right)$
and $\overline{f}_{w_{j}}^{\left(t\right)}\left(\mathbf{w}_{j}\right)$
(i.e. \eqref{eq:surrFunc_p}, \eqref{eq:surrFunc_w}) in this context
possesse the property of asymptotic consistency, which is given in
the key lemma below.

\begin{lem}
[Asymptotic consistency of surrogate functions]\label{lem:Asym-consis-of-suFunc}For
each iteration $t=1,2,\ldots$ and each $\mathbf{p}_{j},\mathbf{w}_{j},j=1,2,\ldots J$,
consider a subsequence $\left\{ \mathbf{p}^{\left(t_{i}\right)},\mathbf{w}^{\left(t_{i}\right)}\right\} _{i=1}^{\infty}$
converging to a limit point $\mathbf{p}^{*},\mathbf{w}^{*}$. There
exist uniformly differentiable functions $\hat{f}_{p_{j}}\left(\mathbf{p}_{j}\right)$
and $\hat{f}_{w_{j}}\left(\mathbf{w}_{j}\right)$ such that
\begin{equation}
\mathop{\lim}\limits _{i\to\infty}\overline{f}_{p_{j}}^{\left(t_{i}\right)}\left(\mathbf{p}_{j}\right)=\hat{f}_{p_{j}}\left(\mathbf{p}_{j}\right),\forall\mathbf{p}_{j},\label{eq:lemma3-1-x}
\end{equation}
\begin{equation}
\mathop{\lim}\limits _{i\to\infty}\overline{f}_{w_{j}}^{\left(t_{i}\right)}\left(\mathbf{w}_{j}\right)=\hat{f}_{w_{j}}\left(\mathbf{w}_{j}\right),\forall\mathbf{w}_{j}.\label{eq:lemma3-1-y}
\end{equation}
Moreover, we have
\begin{equation}
\left\Vert \nabla_{\mathbf{p}_{j}}\hat{f}_{p_{j}}\left(\mathbf{p}_{j}^{*}\right)-\nabla_{\mathbf{p}_{j}}L\left(\mathbf{p}^{*},\mathbf{w}^{*}\right)\right\Vert =0,\label{eq:lemma3-2-x}
\end{equation}
\begin{align}
\left\Vert \nabla_{\mathbf{w}_{j}}\hat{f}_{w_{j}}\left(\mathbf{w}_{j}^{*}\right)-\nabla_{\mathbf{w}_{j}}L\left(\mathbf{p}^{*},\mathbf{w}^{*}\right)\right\Vert  & =0.\label{eq:lemma3-2-y}
\end{align}
\end{lem}

Please refer to Appendix \ref{subsec:Proof-of-Lemma} for the proof.
With the Lemma \ref{lem:Asym-consis-of-suFunc}, we are ready to prove
the following main convergence result.
\begin{thm}
[Convergence of PSPVBI]\label{thm:Convergence-of-SPVBI}Starting
from a feasible initial point $\left\{ \mathbf{p}^{\left(0\right)},\mathbf{w}^{\left(0\right)}\right\} $,
let $\left\{ \mathbf{p}^{\left(t\right)},\mathbf{w}^{\left(t\right)}\right\} _{t=1}^{\infty}$
denote the iterates generated by Algorithm \ref{alg:SPVBI}. Then
every limiting point $\mathbf{p}^{*},\mathbf{w}^{*}$ of $\left\{ \mathbf{p}^{\left(t\right)},\mathbf{w}^{\left(t\right)}\right\} _{t=1}^{\infty}$
is a stationary point of optimization problem $\mathcal{P}_{2}$ almost
surely.
\end{thm}

The proof is similar to that of (\cite{SPVBI_arxiv}, Theorem 3) and
is omitted for conciseness.

\subsection{Convergence analysis of deep-unfolded LPSPVBI}

For deep-unfolding networks, if the iterated algorithm being unfolded
is convergent and the network structure is appropriate, it can typically
achieve convergence after effective training. However, it is diffficult
to provide a rigorous proof of convergence for the deep-unfolded LPSPVBI.
Therefore, in this section, we will intuitively explain the reason
why such an approach works.

Firstly, let us investigate the connection between the deep-unfolded
LPSPVBI and the PSSCA-based PSPVBI algorithm. it is obvious that the
LPSPVBI has a similar structure with the parallel SSCA. However, the
LPSPVBI introduces learnable parameters to accelerate convergence.
Thus, LPSPVBI inherits almost all of PSPVBI's convergence properties,
without altering the original algorithm's working mechanism.

Secondly, proposed algorithm essentially searches for a particle-based
variational distribution in the density space that best fits the true
posteriori, based on a gradient descent strategy. As pointed out in
\cite{Wei_Yi}, any gradient descent-type algorithm selects a direction
$\boldsymbol{d}^{\left(i\right)}$, then searches along this direction
with an appropriate step size $\eta^{\left(i\right)}$ to obtain a
new iteration that achieves a smaller distance to the optimal solution
$\boldsymbol{s}^{*}$. Suppose after $I$ iterations, we can find
the optimal solution, i.e.,
\begin{equation}
\boldsymbol{s}^{*}=\hat{\boldsymbol{s}}^{\left(0\right)}+\eta^{\left(0\right)}\boldsymbol{d}^{\left(0\right)}+\ldots+\eta^{\left(I-1\right)}\boldsymbol{d}^{\left(I-1\right)}.
\end{equation}
The above process can be seen as finding a set of vectors (directions)
and corresponding weights (step sizes) whose weighted linear combination
equals $\boldsymbol{s}^{*}-\hat{\boldsymbol{s}}^{\left(0\right)}$.
In the PSPVBI algorithm, the step sizes are fixed, requiring a larger
number of iterations to approximate the optimal solution. However,
the LPSPVBI algorithm tries to find a more optimal set of stepsizes,
such that the number of iterations can be restricted to the depth
of the network. In addition, due to the existence of non-linear operator
(i.e. projection) and smoothing operator, the step sizes can also
affect the descent direction.

Therefore, the proposed LPSPVBI algorithm can achieve convergence
in a given number of layers with a high probability.

\section{Applications}

As mentioned earlier, deriving Bayesian inference-based algorithms
one by one for each model is a tedious task \cite{BlackboxVI}. However,
our proposed LPSPVBI algorithm framework effectively addresses this
issue. Based on \eqref{eq:gx} and \eqref{eq:gy}, all that is required
is to provide prior information, model-dependent likelihood function
and training data, to accomplish variational approximation of the
posteriori distribution and parameter fine-tuning. Its strong generality
allows practitioners to rapidly design, apply, and modify their data
models without the need to laboriously derive and debug every time.

We shall apply the proposed LPSPVBI to solve the two application problems
in Section \ref{sec:Problem-Formulations}.

\subsection{Example 1 (Wireless cooperative localization)}

Consider the wireless cooperative localization problem in Example
\ref{exa:WL_exam}. The logarithmic likelihood function and logarithmic
prior distribution for this example are
\begin{equation}
lnp\left(\boldsymbol{s}_{i}\right)=\frac{1}{2}ln\frac{\left|\boldsymbol{U}_{i}\right|}{2\pi}+\left[-\frac{1}{2}\left(\boldsymbol{s}_{i}-\boldsymbol{\mu}_{i}\right)^{T}\boldsymbol{U}_{i}\left(\boldsymbol{s}_{i}-\boldsymbol{\mu}_{i}\right)\right],
\end{equation}
\begin{equation}
lnp\left(\boldsymbol{z}|\boldsymbol{s}_{0},\boldsymbol{c}\right)=\frac{N_{R}}{2}ln\frac{\left|U_{ni}\right|}{2\pi}+\underset{i=1}{\overset{N_{R}}{\sum}}\left[-\frac{U_{ni}}{2}\left(\boldsymbol{z}_{i}-h\left(\boldsymbol{s}_{i},\boldsymbol{s}_{0}\right)\right)^{2}\right].
\end{equation}
Additionally, the first derivatives are
\begin{equation}
\frac{\partial lnp\left(\boldsymbol{s}_{i}\right)}{\partial\boldsymbol{s}_{i}}=-\boldsymbol{U}_{i}\left(\boldsymbol{s}_{i}-\boldsymbol{\mu}_{i}\right),\forall i=0,1,\ldots N_{R},
\end{equation}
\begin{equation}
\frac{\partial lnp\left(\boldsymbol{z}|\boldsymbol{s}_{0},\boldsymbol{c}\right)}{\partial\boldsymbol{s}_{0}}=\underset{i=1}{\overset{N_{R}}{\sum}}\left[U_{ni}\left(\boldsymbol{z}_{i}-h\left(\boldsymbol{s}_{i},\boldsymbol{s}_{0}\right)\right)\frac{\frac{10\lambda}{ln10}\left(\boldsymbol{s}_{i}-\boldsymbol{s}_{0}\right)}{\left\Vert \boldsymbol{s}_{i}-\boldsymbol{s}_{0}\right\Vert _{2}^{2}}\right].
\end{equation}

\subsubsection{Performance Comparison}

We use a similar simulation configuration as that in \cite{WL_PASS}.
There are $N_{R}=6$ reference nodes and a target node in a RSS-based
localization scenario, as shown in Fig. \ref{WL}. The precision of
each node initial location is assumed to be $\boldsymbol{U}_{i}=1/10\mathbf{I},\forall i=0,1,\ldots N_{R}$,
where $\mathbf{I}$ denotes the identity matrix. We assume in all
simulations that, $\lambda=3$ and $\phi_{ref}=-5dBm$. In addition,
we assume $r_{s}=50[m]$, $\varepsilon_{i}=4/75$.

\begin{figure}[htbp]
\begin{centering}
\textsf{\includegraphics[scale=0.5]{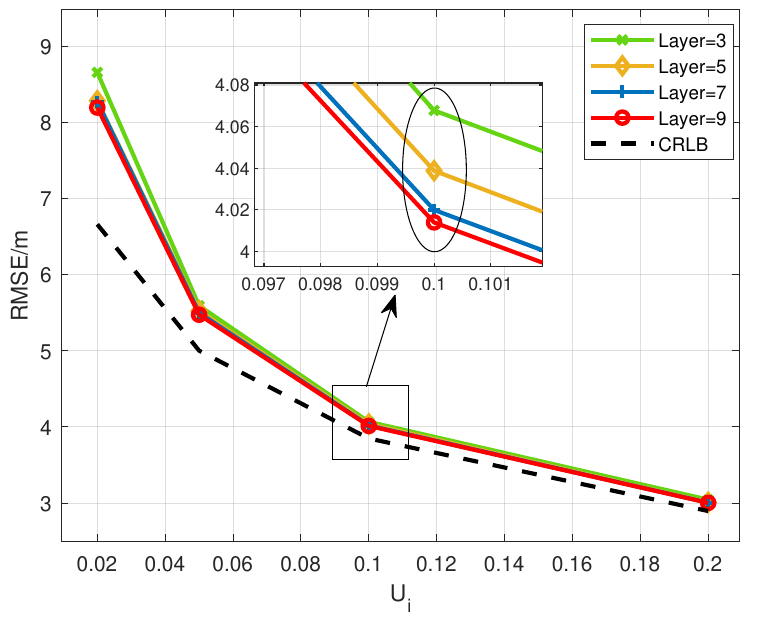}}
\par\end{centering}
\caption{\textsf{\label{diffLayer_U_case1}}Localization error of different
layers over various priori location precisions for LPSPVBI.}
\end{figure}
In Fig. \ref{diffLayer_U_case1}, We plotted the localization error
for different depths under various prior precisions. In general, with
more accurate priors, the positioning error becomes smaller, and the
gap with the Cramer-Rao Lower Bound (CRLB) also decreases. Furthermore,
as we increase the number of network layers from $3$ to $9$, localization
accuracy gradually improves and almost ceases to improve after reaching
a certain number of layers, such as $7$ layers.

\begin{figure}[htbp]
\begin{centering}
\textsf{\includegraphics[scale=0.5]{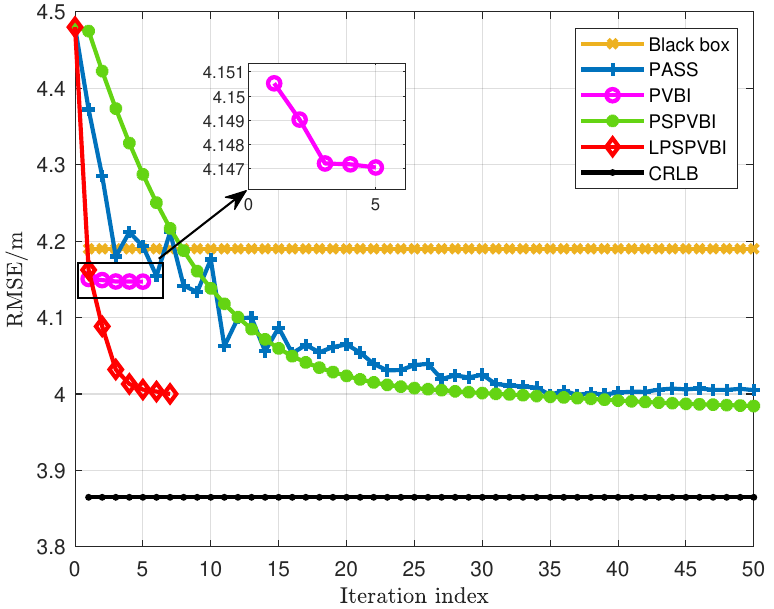}}
\par\end{centering}
\caption{\textsf{\label{iter_WL}}RSS-based localization errors with different
methods.}
\end{figure}
Then, in Fig. \ref{iter_WL}, we compare the performance and convergence
of the proposed PSPVBI, LPSPVBI, and other algorithms. \textbf{1)
Blackbox}: To compare with the dual-driven (i.e. model and data) DU
network, we introduce conventional black-box based DNNs as a benchmark.
Its input is a vector composed of measurements and prior information
about nodes, which sequentially passes through fully connected (FC)
layers, batch normalization (BN), and non-linear functions. The DNN's
output is the position of the target node. We use leaky ReLU as the
activation function and the Adam optimization method. \textbf{2) PASS}
\cite{WL_PASS}: As a particle-based stochastic search algorithm,
PASS lacks theoretical convergence guarantees and heuristically drives
other particles towards the one with the maximum approximate posteriori
probability. \textbf{3) PVBI} \cite{VBI+IS}: PVBI did not update
particle positions and directly computed the multiple summations without
using stochastic approximations. A total of $1000$ experiments were
repeated to obtain the averaged iteration curves shown as in Fig.
\ref{iter_WL}. It can be seen that the LPSPVBI achieves performance
similar to PSPVBI and PASS in very few iterations, demonstrating that
DU effectively accelerates the algorithm's convergence.

\subsubsection{Computational Complexity\label{subsec:exam1_Complexity}}

For PVBI algorithm, its per-iteration complexity order is $\mathcal{O}\left(J\cdot\left(N_{p}\right)^{4}F_{LH}\right)$,
where $F_{LH}$ represents the average number of floating point operations
(FLOPs) required to compute the dominant logarithmic likelihood value.
For PSPVBI algorithm, through mini-batch sampling and minimization
of quadratic surrogate objective functions, the per-iteration complexity
can be reduced to $\mathcal{O}\left(2J\cdot\left(N_{p}BF_{grad}+N_{p}^{3}\right)\right)$,
where $F_{grad}$ and $N_{p}^{3}$ represent the complexity of computing
a gradient and projection, respectively. For LPSPVBI algorithm, the
main complexity lies in the gradient calculation module, and the complexity
of other modules can be neglected. For PASS algorithm, $N_{R}$ represents
the number of reference nodes, while $N_{S}$, $N_{D}$ and $N_{M}$
represent the counts of search particles, detection particles, and
proposal particles respectively. $F_{belief}$ denotes the number
of FLOPs required to compute a single belief. For black-box based
DNN, $L_{0\sim3}$ are the sizes of each layer in DNN. $T_{1\sim4}$
are the typical number of iterations of each algorithm.

In Table \ref{Order_PerIteration_case1}, we summarize the complexity
order of different algorithms, and numerically compare them under
a typical setting as follows: $T_{1}=3$, $J=2$, $N_{R}=6$, $N_{p}=10$,
$F_{LH}=60$; $T_{2}=35$, $N_{S}=20$, $N_{D}=6$, $N_{M}=5$, $F_{belief}=10$;
$T_{3}=25$, $B=20$, $F_{grad}=36$; $L_{1}=22$, $L_{2}=L_{3}=512$,
$L_{4}=2$; $T_{4}=7$. It can be seen that, compared to other algorithms,
LPSPVBI greatly reduces the computational complexity while maintaining
performance.\textcolor{blue}{}
\begin{table}[tbh]
\caption{\label{Order_PerIteration_case1}COMPARISON OF THE COMPLEXITY ORDER
FOR DIFFERENT ALGORITHMS IN EXAMPLE 1}

\centering{}%
\begin{tabular}{|c|c|c|}
\hline
Algorithms & Complexity order & Typical values\tabularnewline
\hline
PVBI & $\mathcal{O}\left(T_{1}J\cdot\left(N_{p}\right)^{4}F_{LH}\right)$ & $36\times10^{5}$\tabularnewline
\hline
PASS & $\mathcal{O}\left(T_{2}N_{R}N_{S}N_{D}N_{M}F_{belief}\right)$ & $12.6\times10^{5}$\tabularnewline
\hline
PSPVBI & $\mathcal{O}\left(T_{3}2J\cdot\left(N_{p}BF_{grad}+N_{p}^{3}\right)\right)$ & $8.20\times10^{5}$\tabularnewline
\hline
Black-box & $\mathcal{O}\left(\sum_{i=0}^{2}L_{i+1}L_{i}\right)$ & $2.74\times10^{5}$\tabularnewline
\hline
LPSPVBI & $\mathcal{O}\left(T_{4}2J\cdot\left(N_{p}BF_{grad}+N_{p}^{3}\right)\right)$ & $2.30\times10^{5}$\tabularnewline
\hline
\end{tabular}
\end{table}

\subsection{Example 2 (Multiband WiFi sensing)}

Consider a multi-band WiFi sensing system in Example \ref{exa:MB_exam},
where WiFi signals from multiple non-contiguous frequency bands are
utilized for ranging purposes. In this scenario, explicit expressions
of the gradient $\nabla_{\mathbf{p}_{j}}g_{j}^{\left(t,b\right)}$
and $\nabla_{\mathbf{w}_{j}}g_{j}^{\left(t,b\right)}$ in \eqref{eq:gx}
and \eqref{eq:gy} are given in the Appendix of \cite{SPVBI_arxiv}.

\subsubsection{Performance Comparison:}

In the simulations, each variable is equipped with $N_{p}=10$ particles,
and the size of mini-batch $B$ is $10$. The received signals come
from two non-adjacent frequency bands with a bandwidth of $20$MHz.
The initial frequency is set to $2.4$GHz and $2.46$GHz, respectively.
The subcarrier spacing is $78.125$KHz. There are two scattering paths
with delays uniformly generated between $20$ns and $200$ns. The
amplitudes $\alpha_{k}$ are $1$ and $0.5$, and the phases $\beta_{k}$
are $-\pi/4$ and $\pi/4$, respectively. In addition, initial phase
$\phi_{m}$ are uniformly generated within $[0,2\pi]$. The timing
synchronization error $\delta_{m}$ is generated following a Gaussian
distribution $\mathcal{N}\left(0,0.01\text{ns}^{2}\right)$. The step
size sequence is set as follows: $\rho^{\left(t\right)}=5/\left(5+t\right)^{0.9},\rho^{\left(0\right)}=1$;
$\gamma^{\left(t\right)}=5/\left(15+t\right)^{1},\gamma^{\left(0\right)}=1$.
Unless otherwise specified, the experiment was repeated $400$ times.

\begin{figure}[htbp]
\begin{centering}
\textsf{\includegraphics[scale=0.5]{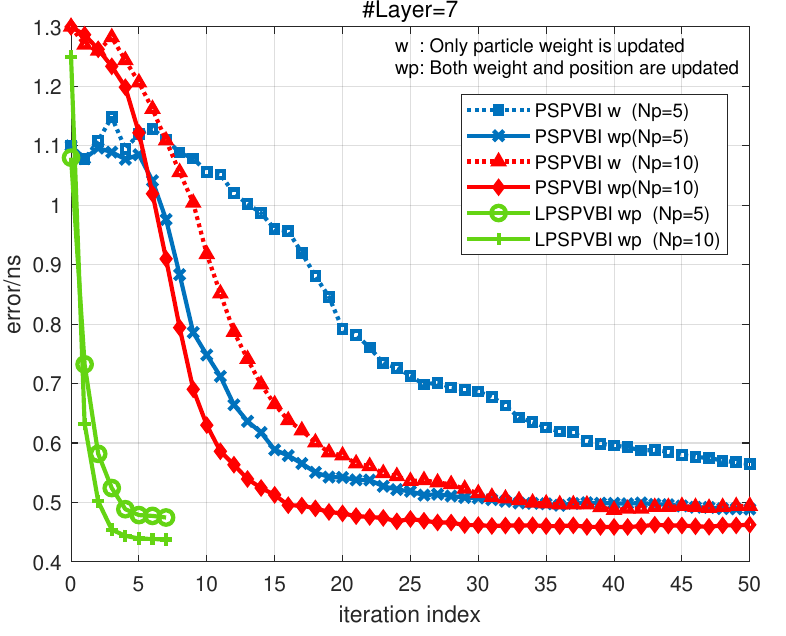}}
\par\end{centering}
\caption{\textsf{\label{diffNumParti_case2}}Iteration error curves with different
particle numbers.}
\end{figure}
In Fig. \ref{diffNumParti_case2}, we plot the convergence curves
of different updating modes with different number of particles. As
the number of particles decreases from $10$ to $5$, it can be seen
that the convergence speed will slow down when only the weight is
updated, but the convergence speed and performance will almost remain
the same when the position is also updated. This means that updating
the position of particles can effectively increase the degree of freedom
of optimization and ensure convergence and performance even with fewer
particles.

\begin{figure}[htbp]
\begin{centering}
\textsf{\includegraphics[scale=0.5]{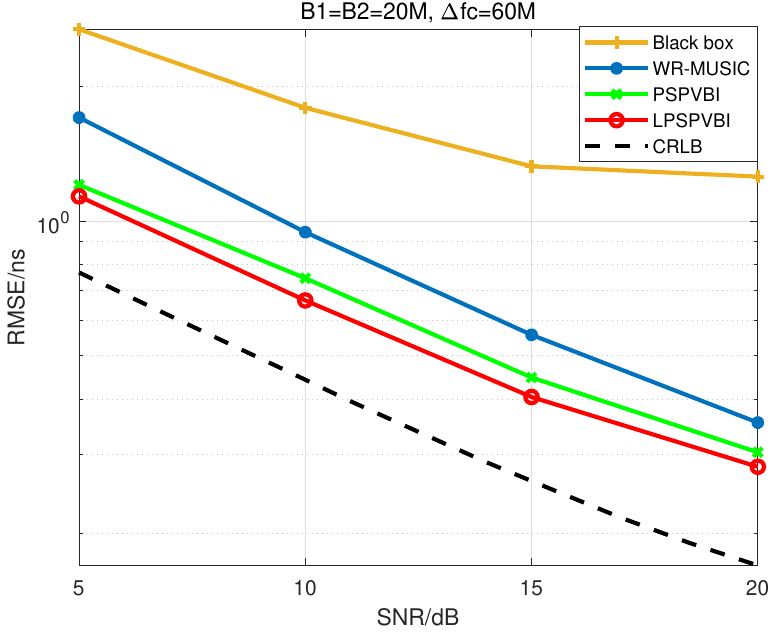}}
\par\end{centering}
\caption{\textsf{\label{diffLayer_SNR_case2}}RMSE of delay estimation with
respect to the SNR.}
\end{figure}

In Fig. \ref{diffLayer_SNR_case2}, we compare the proposed algorithm
with the following baseline algorithms: \textbf{1) Weighted Root-MUSIC
(WR-MUSIC)} \cite{WR-MUSIC}: As an improved subspace-based algorithm,
WR-MUSIC can serve as a leading algorithm to provide prior information
for the PSPVBI. Please refer to \cite{SPVBI_arxiv} for specific details.
\textbf{2) Blackbox}: The structure of black-box based DNN is similar
to that in Example 1. It can be seen that the performance of all algorithms
improves as the signal-to-noise ratio (SNR) increases from $5$dB
to $20$dB. Furthermore, under the adjustment of the Hypernet, the
LPSPVBI algorithm can adapt well to varying SNR conditions. At a certain
SNR, the root-mean-square error (RMSE) of LPSPVBI is closer to the
CRLB and significantly lower than that of other algorithms, indicating
higher time delay estimation accuracy. In addition, compared to Example
1, Example 2 has a higher-dimensional variable space and more local
optima, making the superiority of the proposed algorithm more pronounced.

\subsubsection{Computational Complexity:}

For WR-MUSIC algorithm, it mainly includes subspace decomposition
and polynomial rooting, with the complexity of $\mathcal{O}\left(MN_{m}^{3}\right)$
and $\mathcal{O}\left(M\left(2N_{m}-1\right)^{3}\right)$. The complexity
analysis of other algorithms is similar to Section \ref{subsec:exam1_Complexity}.

In Table \ref{Order_PerIteration_case2}, we summarize the complexity
order of different algorithms, and numerically compare them under
a typical setting as follows: $M=2$, $N_{m}=256$; $L_{0}=1028$,
$L_{1}=3072$, $L_{2}=4096$, $L_{3}=2048$, $L_{4}=256$, $L_{5}=2$;
$T_{1}=3$, $J=9$, $N_{p}=10$, $F_{LH}=5000$, $\left|\Omega\right|=64$;
$T_{2}=35$, $T_{3}=7$, $B=10$, $F_{grad}=4500$.\textcolor{blue}{}
\begin{table}[tbh]
\caption{\label{Order_PerIteration_case2}COMPARISON OF THE COMPLEXITY ORDER
FOR DIFFERENT ALGORITHMS IN EXAMPLE 2}

\centering{}%
\begin{tabular}{|c|c|c|}
\hline
Algorithms & Complexity order & Typical values\tabularnewline
\hline
PVBI\tablefootnote{Due to the large number of variables in this scenario, PVBI algorithm
with excessive computational complexity is not applicable. The performance
comparison between PVBI and PSPVBI in a simplified scenario can be
found in \cite{SPVBI_arxiv}.} & $\mathcal{O}\left(T_{1}J\cdot\left(N_{p}\right)^{J}F_{LH}\right)$ & $1.35\times10^{14}$\tabularnewline
\hline
WR-MUSIC & $\mathcal{O}\left(MN_{m}^{3}+M\left(2N_{m}-2\right)^{3}\right)$ & $2.99\times10^{8}$\tabularnewline
\hline
PSPVBI & $\mathcal{O}\left(T_{2}2J\cdot\left(N_{p}BF_{grad}+N_{p}^{3}\right)\right)$ & $2.84\times10^{8}$\tabularnewline
\hline
LPSPVBI & $\mathcal{O}\left(T_{3}2J\cdot\left(N_{p}BF_{grad}+N_{p}^{3}\right)\right)$ & $5.68\times10^{7}$\tabularnewline
\hline
Black-box & $\mathcal{O}\left(\sum_{i=0}^{4}L_{i+1}L_{i}\right)$ & $2.47\times10^{7}$\tabularnewline
\hline
\end{tabular}
\end{table}

By comparing the performance and complexity of the algorithms before
and after unfolding, we can conclude that deep-unfolding indeed fully
unleashes the algorithm's potential.

\section{Conclusions}

In this paper, we consider a general parameter estimation problem
in sensing scenarios. Within the framework of Bayesian estimation,
parameter estimation is transformed into a variational optimization
problem aimed at approximating the true posteriori distribution. To
address it, we propose a parallel stochastic particle variational
Bayesian inference (PSPVBI) algorithm to find a stationary point of
this problem. Compared to conventional VBI, particle approximation
and particle position update make it more flexible and faster in fitting
the posteriori distribution. Furthermore, the introduction of parallel
stochastic successive convex approximation (PSSCA) also effectively
addresses the pain point of high computational complexity in variational
inference. To further reduce complexity, we perform a deep-unfolding
for PSPVBI algorithm by constructing a model-driven deep learning
network. The derived learnable PSPVBI (LPSPVBI) algorithm can further
decrease the number of iterations and single-iteration complexity.
Some techniques related to subsampling, gradient computation, and
generalization ability used in this approach also contribute to the
practical deployment of the algorithm. Finally, we use several important
application examples to demonstrate the effectiveness of the proposed
algorithm. The low-complexity LPSPVBI algorithm holds the potential
to be applied in a broad range of high-dimensional, non-convex parameter
estimation scenarios, even with numerous local optima and non-conjugate
priors.

\appendix

\subsection{Proof of Lemma \ref{lem:Asym-consis-of-suFunc} \label{subsec:Proof-of-Lemma}}

We first introduce the following preliminary result.
\begin{lem}
\label{lem:Properties-of-surrFunc}Given problem $\mathcal{P}_{2}$,
suppose that the step sizes $\rho^{\left(t\right)}$and $\gamma^{\left(t\right)}$
are chosen according to step-size rules in \ref{subsec:SPVBI-Algorithm-Design}.
Let $\left\{ \mathbf{p}^{\left(t\right)},\mathbf{w}^{\left(t\right)}\right\} $
be the sequence generated by Algorithm \ref{alg:SPVBI}. Then, the
following holds
\begin{equation}
\mathop{\lim}\limits _{t\to\infty}\left|\mathbf{\boldsymbol{{\rm f}}}_{p_{j}}^{\left(t\right)}-\nabla_{\mathbf{p}_{j}}\boldsymbol{L}\left(\mathbf{p}^{\left(t\right)},\mathbf{w}^{\left(t\right)}\right)\right|=0,w.p.1.
\end{equation}
\begin{equation}
\mathop{\lim}\limits _{t\to\infty}\left|\mathbf{\boldsymbol{{\rm f}}}_{w_{j}}^{\left(t\right)}-\nabla_{\mathbf{w}_{j}}\boldsymbol{L}\left(\mathbf{p}^{\left(t\right)},\mathbf{w}^{\left(t\right)}\right)\right|=0.w.p.1.
\end{equation}
\end{lem}
\begin{IEEEproof}
Lemma \ref{lem:Properties-of-surrFunc} is a consequence of (\cite{lemma},
Lemma 1). We only need to verify that all the technical conditions
(a)\textendash (e) therein are satisfied. Specifically, Condition
(a) of (\cite{lemma}, Lemma 1) is satisfied because domains of particle
positions $\mathbf{p}$ and weights $\mathbf{w}$ are compact and
bounded. Furthermore, the boundedness of $\mathbf{p}$ and $\mathbf{w}$
ensures that the logarithmic terms in \eqref{eq:gx} and \eqref{eq:gy}
do not tend towards infinity. As a result, the instantaneous gradient
$\nabla g_{j}^{\left(t\right)}$ (i.e. $\nabla_{\mathbf{p}_{j}/\mathbf{w}_{j}}g_{j}^{\left(t\right)}$)
remains bounded, satisfying Condition (b) of (\cite{lemma}, Lemma
1). Conditions (c)\textendash (d) directly derived from the step-size
rule 1) in \ref{subsec:SPVBI-Algorithm-Design}. When the particles
of all variables are updated in parallel, the distributions (i.e.,
positions and weights of particles) are no longer identical over iterations,
but change slowly at the rate of order $\mathcal{O}\left(\gamma^{\left(t\right)}\right)$,
which causes the random state $\boldsymbol{\theta}$ in the objective
function to change constantly. Due to the boundedness of $\mathbf{p}$
and $\mathbf{w}$, all the logarithmic terms in the instantaneous
gradient $\nabla g_{j}^{\left(t\right)}$ are Lipschitz continuous.
Therefore, $\nabla g_{j}^{\left(t\right)}$ is also Lipschitz continuous,
and $\left\Vert \nabla g_{j}^{\left(t+1\right)}-\nabla g_{j}^{\left(t\right)}\right\Vert =\mathcal{O}\left(\gamma^{\left(t\right)}\right)$.
Additionally, the expectation operator (i.e. the multiplication and
addition operations with $\mathbf{w}$) slows down the change of $\nabla\boldsymbol{L}_{j}^{\left(t\right)}$
(i.e. $\nabla_{\mathbf{p}_{j}/\mathbf{w}_{j}}\boldsymbol{L}^{\left(t\right)}$),
resulting in $\left\Vert \nabla\boldsymbol{L}_{j}^{\left(t+1\right)}-\nabla\boldsymbol{L}_{j}^{\left(t\right)}\right\Vert =\mathcal{O}^{2}\left(\gamma^{\left(t\right)}\right)$.
Plusing the step-size rule 2) in \ref{subsec:SPVBI-Algorithm-Design},
Condition (e) of (\cite{lemma}, Lemma 1) $\nicefrac{\left\Vert \nabla\boldsymbol{L}_{j}^{\left(t+1\right)}-\nabla\boldsymbol{L}_{j}^{\left(t\right)}\right\Vert }{\rho^{\left(t\right)}\rightarrow0}$
is also satisfied.
\end{IEEEproof}
As can be seen, Lemma \ref{lem:Asym-consis-of-suFunc} requires proof
for two parts. Similar to the proof in (\cite{SSCA}, Lemma 1), we
first establish the existence of the limit of the surrogate function
sequence, i.e., \eqref{eq:lemma3-1-x} and \eqref{eq:lemma3-1-y}.

Since the surrogate functions adopted are convex quadratic functions
with box constraints/simplex constraints, it is straightforward to
determine that they possess the following properties. For any $\mathbf{p}_{j}\in\mathcal{X}_{p}$
and $\mathbf{w}_{j}\in\mathcal{\mathcal{X}}_{w}$, $\overline{f}_{p_{j}}^{\left(t\right)}\left(\mathbf{p}_{j}\right)$
and $\overline{f}_{w_{j}}^{\left(t\right)}\left(\mathbf{w}_{j}\right)$
are uniformly strongly convex and Lipschitz continuous, and their
derivative, second order derivative are uniformly bounded \cite{SSCA,SPVBI}.
Due to these properties of the surrogate functions, the families of
functions $\left\{ \overline{f}_{p_{j}}^{\left(t_{i}\right)}\left(\mathbf{p}_{j}\right)\right\} $
and $\left\{ \overline{f}_{w_{j}}^{\left(t_{i}\right)}\left(\mathbf{w}_{j}\right)\right\} $
are equicontinuous. Moreover, they are bounded and defined over a
compact set $\mathcal{X}_{p}$ and $\mathcal{\mathcal{X}}_{w}$. Hence
the Arzela\textendash Ascoli theorem \cite{1988Linear} implies that,
by restricting to a subsequence, there exists uniformly continuous
functions $\hat{f}_{p_{j}}\left(\mathbf{p}_{j}\right)$ and $\hat{f}_{w_{j}}\left(\mathbf{w}_{j}\right)$
such that \eqref{eq:lemma3-1-x} and \eqref{eq:lemma3-1-y} in Lemma
\ref{lem:Asym-consis-of-suFunc} are satisfied.

Next, we will demonstrate the consistency of gradients of the surrogate
function at the limit point. Clearly, according to the quadratic surrogate
functions in \eqref{eq:surrFunc_p} and \eqref{eq:surrFunc_w}, we
have $\mathop{\lim}\limits _{i\to\infty}\mathbf{\boldsymbol{{\rm f}}}_{p_{j}}^{\left(t_{i}\right)}=\mathop{\lim}\limits _{i\to\infty}\nabla_{\mathbf{p}_{j}}\overline{f}_{p_{j}}^{\left(t_{i}\right)}\left(\mathbf{p}_{j}^{\left(t_{i}\right)}\right)$
and $\mathop{\lim}\limits _{i\to\infty}\mathbf{\boldsymbol{{\rm f}}}_{w_{j}}^{\left(t_{i}\right)}=\mathop{\lim}\limits _{i\to\infty}\overline{f}_{w_{j}}^{\left(t_{i}\right)}\left(\mathbf{w}_{j}^{\left(t_{i}\right)}\right)$.
And because of \eqref{eq:lemma3-1-x} and \eqref{eq:lemma3-1-y},
we further have $\mathop{\lim}\limits _{i\to\infty}\mathbf{\boldsymbol{{\rm f}}}_{p_{j}}^{\left(t_{i}\right)}=\nabla_{\mathbf{p}_{j}}\hat{f}_{p_{j}}\left(\mathbf{p}_{j}^{*}\right)$
and $\mathop{\lim}\limits _{i\to\infty}\mathbf{\boldsymbol{{\rm f}}}_{w_{j}}^{\left(t_{i}\right)}=\nabla_{\mathbf{w}_{j}}\hat{f}_{w_{j}}\left(\mathbf{w}_{j}^{*}\right)$.
Finally, using Lemma \ref{lem:Properties-of-surrFunc}, we have
\begin{align}
\left\Vert \nabla_{\mathbf{p}_{j}}\hat{f}_{p_{j}}\left(\mathbf{p}_{j}^{*}\right)-\nabla_{\mathbf{p}_{j}}L\left(\mathbf{p}^{*},\mathbf{w}^{*}\right)\right\Vert  & =0,\\
\left\Vert \nabla_{\mathbf{w}_{j}}\hat{f}_{w_{j}}\left(\mathbf{w}_{j}^{*}\right)-\nabla_{\mathbf{w}_{j}}L\left(\mathbf{p}^{*},\mathbf{w}^{*}\right)\right\Vert  & =0,
\end{align}
almost surely.

\bibliographystyle{IEEEtran}
\bibliography{LSP-VBI}

\end{document}